\documentclass[twocolumn]{aastex631}

\usepackage[utf8]{inputenc}
\usepackage{graphicx}
\graphicspath{{figures/}}
\usepackage[version=4]{mhchem}
\usepackage{booktabs}
\usepackage{bm}
\usepackage{placeins}

\usepackage{tabularx}
\usepackage{xspace}
\usepackage{threeparttablex}
\usepackage{ragged2e}
\newcolumntype{Y}{>{\RaggedRight\arraybackslash}X}

\begin{document}

\title{Rotational Mapping and Regional Atmospheric Retrievals of Variable Brown Dwarfs:\\ Application to Luhman 16B and SIMP 0136}

\author[0009-0007-8482-2704]{Ruizhe Wang}
\email{rwang3@caltech.edu}
\affiliation{Division of Geological and Planetary Sciences, California Institute of Technology, Pasadena, CA 91125, USA}

\author[0000-0002-5375-4725]{Heather A. Knutson}
\affiliation{Division of Geological and Planetary Sciences, California Institute of Technology, Pasadena, CA 91125, USA}

\author[0000-0001-9164-7966]{Julie Inglis}
\affiliation{Division of Geological and Planetary Sciences, California Institute of Technology, Pasadena, CA 91125, USA}
\affiliation{Department of Astronomy and Astrophysics, University of California, San Diego, La Jolla, CA 92093, USA}

\author[0000-0003-2278-6932]{Xianyu Tan}
\affiliation{Tsung-Dao Lee Institute \& School of Physics and Astronomy, Shanghai Jiao Tong University, Shanghai 201210, China}

\begin{abstract}
We present a Bayesian framework for converting time-resolved spectroscopy into rotation-inferred regional spectra and atmospheric constraints for variable brown dwarfs. We derive an analytic mapping framework with evidence-optimized regularization and uncertainty propagation by marginalizing over physical parameters and model complexities. We provide methods for selecting characteristic spectral end-members from the inferred surface maps, and highlight the advantage of analyzing these rotation-inferred regional spectra over the disk-integrated time-resolved observations, while explicitly accounting for the geometric degeneracies inherent to rotational mapping. We apply this framework to published JWST/NIRSpec rotational monitoring of Luhman~16B and SIMP~0136. For both objects, we identify dominant spectral regions and retrieve regional variations in cloud structure, thermal gradients, and chemistry that are consistent with cloud radiative feedback. In addition, we demonstrate the importance of limb darkening in deriving surface maps. These results establish a statistically controlled method for linking rotational spectral variability to atmospheric heterogeneity and enable future rotational mapping and regional atmospheric studies of giant planet atmospheres.
\end{abstract}

\section{Introduction}
\label{sec:introduction}


Many directly imaged planets and brown dwarfs have effective temperatures between $1000$ and $2000$~K \citep{Currie2022}. At these temperatures, silicates and other refractory species condense out of the gas phase and form thick high-altitude clouds that can drastically alter the observed spectral shapes of these objects \citep{Robinson2015}. Published studies of brown dwarfs and directly imaged planets in this temperature range find that they frequently exhibit rotational variability of a few percent or more \citep[e.g.,][]{Zhou2022, Vos2023}, which has been attributed to inhomogeneities in cloud coverage closely tied to atmospheric circulation patterns \citep{Showman2009, Showman2020, Tan2021}.

General circulation models (GCMs) for brown dwarfs predict vigorous motion in these atmospheres \citep[e.g.,][]{Zhang2014, Tan2021, Tan2021tdb, Tan2025, Teinturier2026}. Observational measurements also support fast zonal winds in brown dwarf atmospheres \citep{Allers2020}. Rotation, convection, and cloud radiative feedback together shape the weather patterns. Because brown dwarfs are internally heated by convection from the deep interior, turbulent transport of energy injects eddies and induces waves in the stably stratified layer above the convective interior \citep{Showman2012}. Since energy in turbulent quasi-geostrophic flows tends to grow upscale (inverse energy cascade), the turbulence grows and reorganizes into coherent, large-scale flows \citep[e.g.,][]{Tan2021,Tan2025}. Notably, current GCMs of brown dwarfs predict few broad jets in the atmosphere \citep{Tan2021}, in contrast to Jupiter's structure of narrow alternating bands of winds \citep[e.g.,][]{Ingersoll2004}.


Clouds add a layer of complexity to this picture, as they prevent thermal radiation from escaping to space; cloudy regions therefore experience different vertical profiles of heating and cooling compared to clear regions \citep{Tan2021tdb}. This process, known as cloud radiative feedback, leads to horizontal temperature differences on isobars, which can drive circulation that transports cloud condensates vertically, maintaining and triggering cloud cycles. \cite{Lee2023} showed that cloud radiative feedback can also alter local atmospheric chemistry through sustained temperature differences and chemical transport.

The resulting spatial heterogeneity is predicted to manifest as rotational variability in the observed spectra, but there is no definitive consensus in the literature as to whether the observed infrared variability of brown dwarfs at these temperatures is primarily driven by spatial variations in cloud properties, thermal structures, chemical abundances, or a combination of these effects \citep[e.g.,][]{McCarthy2025, Nasedkin2025, Oliveros-Gomez2025, Teinturier2026, Schrader2026}.

With the advent of the James Webb Space Telescope (JWST), spectroscopic monitoring campaigns of several benchmark brown dwarfs and directly imaged planets have now made it possible to test the predictions of atmospheric circulation models by exploring the spatially varying properties of these objects in unprecedented detail. For example, \cite{Biller2024} presented time-resolved spectra for Luhman 16AB observed with NIRSpec/PRISM and MIRI/LRS. \cite{McCarthy2025} presented observations for SIMP 0136 in the same settings. Both studies used principal component analysis (PCA) to group light curves from different wavelength bins based on their shapes and link them to absorbing species. They found evidence for clouds, hot spots, and changing carbon chemistry, and concluded that no single mechanism can fully explain the observed variations.

Following up on these observations, \cite{Oliveros-Gomez2025} interpreted the pressure- and molecule-dependent light-curve amplitudes of Luhman 16B using contribution functions and planetary-scale waves; \cite{Nasedkin2025} performed independent atmospheric retrievals on 24 phase-resolved spectra of SIMP 0136 and concluded that variability is more likely caused by temperature differences than by cloud coverage; \cite{Akhmetshyn2025} presented observations of SIMP 0136 with NIRISS/SOSS and compared the time-resolved spectra with self-consistent 1D models. Relevant to our studies, they discussed the difficulties caused by degeneracies in brightness maps inferred from rotational light curves. Using the same dataset, \cite{Wang2026} instead characterized spatial variations in atmospheric properties from principal-component eigenspectra of the disk-integrated time series. Recently, \cite{Schrader2026} used PCA to interpret the time series spectra of SIMP 0136, representing the time series as a mixture of three extreme spectral states. They linked each spectral state with a self-consistent 1D model, showing that changes in temperature and cloud vertical structure account for most of the variability.


These studies provided important insights into the underlying causes of the observed variability, but did not directly infer regional atmospheric properties. Crucially, each time-resolved spectrum is a disk-integrated measurement that contains contributions from an entire hemisphere. This means that any heterogeneity is effectively smeared out by the disk integration, reducing the amplitude of spectral shape variations and making it harder to measure regional differences in atmospheric properties using retrievals. In order to mitigate this issue and leverage the rich information content of these datasets, we introduce a new framework for inverting spectroscopic rotational light curves into regional spectra while accounting for the geometric degeneracies inherent in the mapping process. By grouping the inferred spectra into distinct regions, we can then perform atmospheric retrievals on dominant spectral end-members to identify atmospheric states consistent with the observed variability.

In this study, we develop and apply this framework to these two benchmark cloudy brown dwarfs with extensive JWST/NIRSpec observations. Our first target is the brown dwarf binary Luhman 16AB (WISE J104915.57-531906.1AB), located at a distance of 2 pc \citep{Luhman2013}. Luhman 16B is highly variable ($5$--$15$\%), making it an ideal target for variability mapping. It is viewed nearly equator-on (\(i \gtrsim 80^\circ\)), as inferred from the combination of its rotation period ($P\approx5.3$ hours) and projected rotational velocity \citep{Apai2021}. It was observed with NIRSpec and MIRI during two separate epochs as part of GO 2965 (PI: Biller). In this study, we analyze the 8-hour time-resolved NIRSpec sequence of Luhman 16B observed on 2023 July 8, which spans $\sim$1.6 rotations \citep{Biller2024}. A second full-period observation of Luhman 16B from the same program, obtained on 2024 February 28, was presented by \cite{Chen2025}. Because brown dwarf cloud patterns often evolve on short timescales \citep[e.g.,][]{Tan2025}, the atmospheric structure likely differs between these two epochs. Here, we focus on a single observational epoch to demonstrate our method, and leave a joint analysis of both epochs to explore its long-term atmospheric evolution for future work.

Our second target is SIMP J013656.5+093347.3 (hereafter SIMP 0136), a well-studied variable T2.5 brown dwarf. Its near-infrared variability was first reported by \citet{Artigau2009}, who measured a periodic modulation near 2.4 hr with a peak-to-peak $J$-band amplitude of $\sim50$ mmag. Subsequent Spitzer monitoring refined the rotation period to $P_{\rm rot}=2.414\pm0.078$ hr \citep{Yang2016}; combining this period with a $v\sin i$ measurement gives an inclination of $i=80^{+10}_{-12}$ deg \citep{Vos2017}. Here, we examine the time-resolved NIRSpec observations from GO 3548 \citep[PI: Vos;][]{McCarthy2025}, which span 3.4 hours ($\sim$1.4 rotations). We obtain the reduced time-series spectra for these observations from \cite{Nasedkin2025}. For both objects, the NIRSpec data cover 0.6--5.3 $\mu$m at resolving powers $R\sim30-300$.

\section{Deriving Rotation-Inferred Regional Spectra}
\label{sec:mapping}

There is an extensive literature laying out the mathematical foundation for inverting rotational light curves to produce brightness maps \citep[e.g.,][]{Cowan2013,Luger2019,Luger2021w9k,Luger2021s1,Luger2021h2}. \cite{Luger2019} showed that when we express the surface brightness distribution (surface map) of an object as the sum of a series of spherical harmonics, we can derive an analytic, closed-form solution for the total flux received from the object as a function of rotational phase. This allows us to solve the inverse problem of inferring the surface map from the observed light curve. We build on the analytic mapping framework implemented in the open-source \texttt{Starry} package \citep{Luger2019} to invert the JWST NIRSpec spectroscopic time series for Luhman 16B and SIMP 0136 into wavelength-dependent surface maps. In the following subsections, we describe our method for deriving rotation-inferred surface spectra, including our mapping inversion framework and approach to identifying spectrally distinct regional end-members from these surface maps.

\subsection{Mapping Model Parameterization, Regularization Prior, and Uncertainty Propagation}
\cite{Luger2021h2} explored the degeneracies inherent to light-curve inversion with spherical harmonics and clarified which surface-brightness modes can and cannot be constrained from unresolved photometry. Similarly, \cite{Challener2023NullSpace} discussed complications of the eclipse mapping problem. In both settings, modes that do not affect the observed light curves are in the null space of the linear inverse problem, so infinitely many surface maps can provide statistically equivalent fits to the data. Previous studies have often managed this degeneracy by excluding modes in the null space and restricting the spherical-harmonic expansion to low orders, typically \(l_{\rm max}\sim2\)--4, using the Bayesian Information Criterion or similar model-selection criteria to justify the selected model complexity \citep[e.g.,][]{Rauscher2018, Akhmetshyn2025, Challener2025}. This strategy is useful for finding a compact map that fits the light curve, but it does not characterize the full set of maps that are consistent with the observations. The modes that are not constrained by data should not simply be excluded, but rather regularized using a prior. Excluding those modes artificially narrows the inferred map uncertainties. In our framework, we therefore include these under-constrained modes in the inference and use an explicit, physically motivated prior to regularize their amplitudes.

Here, we consider a broad range of models and use Bayesian model averaging \citep[BMA;][]{Hoeting1999} to marginalize over models of varying complexity (see \S\ref{sec:harmonic_degree} for more details). We additionally adopt a Bayesian framework to propagate the uncertainties in the derived maps to the subsequent analyses, where we include a prior to reflect the fact that we expect there to be relatively less power in the higher-order spherical harmonics. We show that the posterior distribution of surface maps remains closed-form under our regularization.

Many published mapping studies of brown dwarfs and transiting exoplanets either report a best-fit map solution \citep[e.g.,][]{Akhmetshyn2025} or use Monte Carlo fits to numerically determine the uncertainties on the maps \citep[e.g.,][]{Karalidi2015,Rauscher2018}. With our approach, the closed-form nature of the solution facilitates the calculation of the full Bayesian evidence, making it trivial to marginalize over different model parameterizations (e.g., varying spherical harmonic degrees) and different physical parameters (e.g., inclination and rotational period) rather than fixing them. This makes our mapping framework efficient and scalable to arbitrarily many wavelength channels. For more information on this Bayesian mapping framework, see Appendix \ref{sec:Bayesian_Formulation}.

Since the posteriors of the under-constrained modes are dominated by the regularization, we need the regularization to reflect our prior knowledge of the surface. We considered the following factors when choosing this regularization:

\begin{enumerate}
    \item The prior should be conjugate to the Gaussian likelihood, allowing the posterior distribution to remain Gaussian and closed-form. This limits our prior to a multivariate Gaussian function, also known as Tikhonov regularization \citep{Tikhonov1977}, where we specify the mean and covariance.

    \item The strength of the regularization, equivalently the width of the Gaussian prior, should not be chosen arbitrarily. It should either encode a physically motivated expectation for the atmospheric structure or be inferred from the modes that are constrained by the data. For example, \cite{Luger2021s1} derived the mean and covariance of the spherical-harmonic coefficients from a parameterized spot model, in which the prior is specified by quantities such as the number of spots, their characteristic sizes, their latitude distribution, and their contrast. For brown dwarf atmospheres, however, we do not have strong prior constraints on the analogous quantities. One could in principle use a GCM tailored to the object to estimate the angular power spectrum of the brightness map, but such models are expensive to compute and may still omit relevant cloud, chemical, or dynamical processes. We therefore adopt a data-driven prior as the most practical choice.

    A related physically motivated alternative would be to model the surface brightness as a Gaussian process on the sphere, with a covariance kernel specified by a correlation length scale and amplitude, and then project that prior onto the spherical-harmonic coefficients. This approach would make the assumed spatial smoothness explicit, but the GP hyperparameters would still need to be chosen or inferred from the data. We defer exploration of such priors to future studies.

    \item In the case of a data-driven, evidence-optimized prior, we have no reason to expect any particular spatial pattern a priori. Therefore, the prior should be centered on a uniform, constant surface map. This means that the mean of the Gaussian prior should be zero (except for the constant term, which should be one) and the covariance matrix should be diagonal. The variance of each spherical harmonic coefficient reflects the expected power in that mode, which should be object-dependent. For example, we expect the angular power to decrease with increasing angular degree (commonly denoted by $l$), because nature prefers smoothness.
    We analyzed the angular power spectrum of a published brown dwarf GCM \citep{Tan2025} and found that the power in the higher angular degrees decays exponentially after the first few degrees.
    Since we are truncating higher angular degrees from the fit, we apply a constant prior width to all harmonics. This choice reduces Tikhonov regularization to Ridge (L2) regularization \citep{HoerlKennard1970}, which enables optimization of the prior width by iteratively maximizing the model evidence. \cite{mackay1992a} provided the recipe of this procedure and proved the convergence to optimality.

\end{enumerate}

This framework offers several practical advantages over previous approaches. First, it avoids committing to an arbitrary cutoff in the maximum spherical-harmonic degree by marginalizing over a range of model complexities. Second, it determines the strength of the regularization prior by maximizing the model evidence. In practice, this estimates the angular power scale of the underlying map from the data-constrained modes and propagates that scale into the null space. Compared with approaches that sample spherical-harmonic coefficients under fixed, arbitrarily chosen priors, our method provides a more self-consistent estimate of the mapping uncertainties while avoiding the computational cost of Monte Carlo sampling.

Finally, the analytic solution lets us propagate the mapping posterior into the regional spectra, including the covariance induced between regions by the common light-curve inversion. This becomes especially important for JWST-quality observations with hundreds of wavelength channels. The independent atmospheric retrievals in Section~\ref{sec:retrieval}, however, use only each region's marginal variance. A joint fit of all regional spectra can utilize that covariance information.

In this study, we invert surface maps at different wavelengths independently. However, we know a priori that the brightness maps at adjacent pressure layers are coupled. An extension of the framework is to model the surface maps at different wavelengths jointly as a Gaussian process, with a covariance matrix to couple wavelengths that probe nearby pressure layers together. We can then solve wavelength-dependent maps jointly, coupling nearby layers together while keeping the framework analytic. We defer this extension to future studies.


\subsubsection{Marginalization over Spherical Harmonic Degree}\label{sec:harmonic_degree}

When fitting surface maps with spherical harmonics, we must choose the maximum angular degree included in the expansion. A larger \(l_{\rm max}\) allows more spatial structure, but also increases the number of weakly constrained modes and the risk of overfitting. A smaller \(l_{\rm max}\) gives a simpler model, but excludes uncertainty from higher-degree modes. We mitigate this model-selection problem using Bayesian model averaging (BMA), which accounts for structural uncertainty by evaluating a suite of candidate models rather than selecting a single ``best'' model. In our analysis, we evaluate models with \(l_{\rm max}=2,\ldots,10\) and marginalize over this discrete set using the model evidence.

Although this procedure still imposes a finite upper limit on the harmonic degree, \(l_{\rm max}=10\) is matched to the spatial scale of our downstream science product. We do not interpret individual pixels in the fine surface map; instead, we use the maps to construct average spectra of larger spectrally similar regions. We evaluate the spherical-harmonic maps on a fine pixelated grid, identify spectrally similar regions, and average the spectra within each region (Section~\ref{sec:grouping}). The spatial scales that can be meaningfully interpreted in these regional spectra are therefore tens of degrees rather than individual grid cells. Since a spherical harmonic of degree \(l\) has a characteristic angular scale of approximately \(180^\circ/l\), modes with \(l\gg10\) describe structures substantially smaller than the scales retained in the region-averaged spectra. These fine-scale modes largely cancel when averaged over the identified regions, so including them would mainly add small-scale map uncertainty without changing the regional spectra. Thus, truncating the candidate models at \(l_{\rm max}=10\) preserves uncertainty on the spatial scales relevant to our analysis while avoiding unsupported interpretation of fine-scale surface structure.

The final solution is then constructed as the average of all considered models, weighted by each model's predictive power (evidence). This marginalization is straightforward in our framework because the inference problem is linear and has closed-form solutions. See Appendix \ref{sec:Bayesian_Formulation} for the details behind the marginalization.

\begin{figure*}[t!]
\centering
\includegraphics[width=\textwidth]{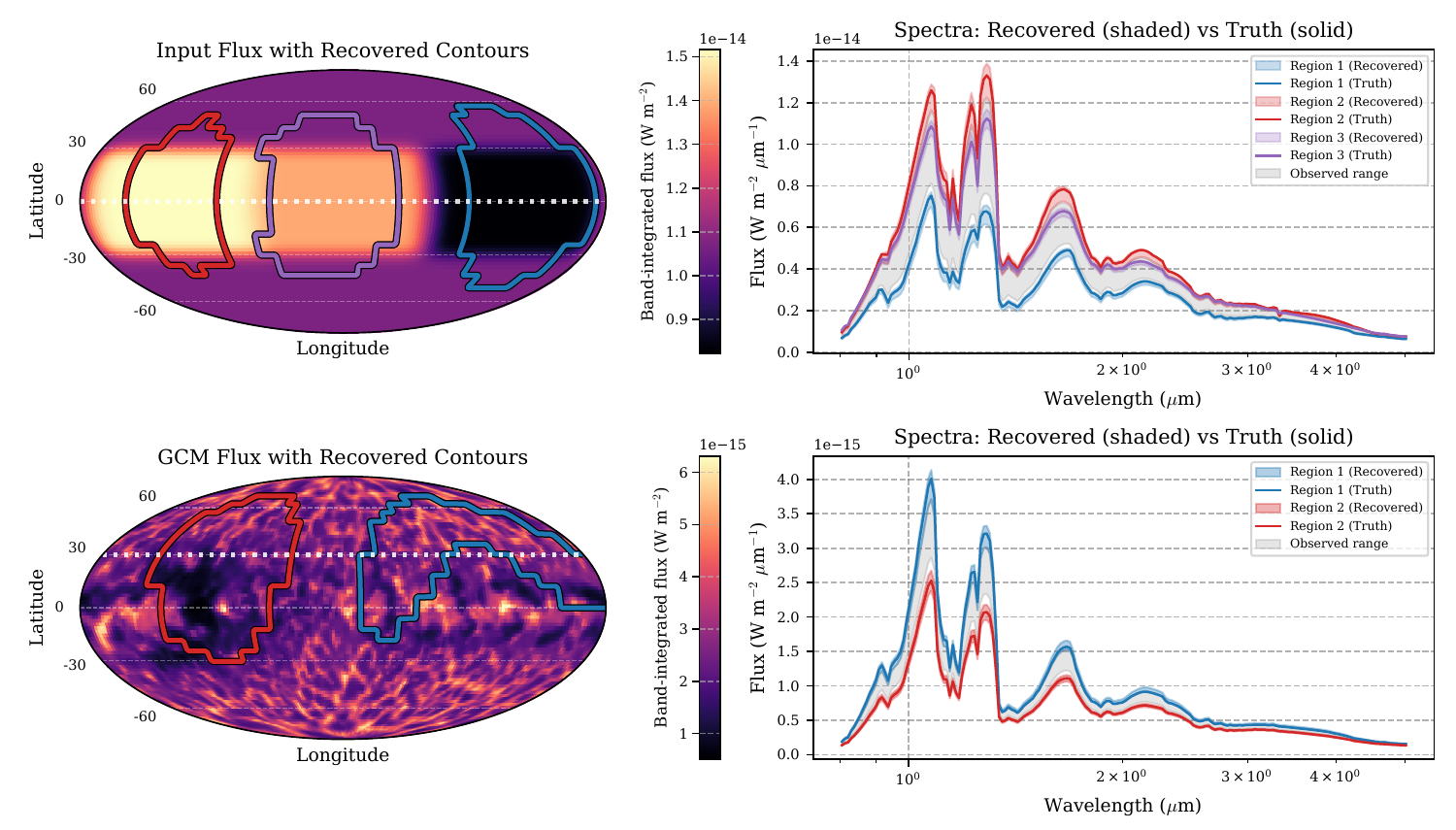}
\caption{Left: comparison of the spatial input (ground truth) versus the recovered regions. In both cases, the contours encircle the identified spectrally distinct regions. The white dotted line denotes the sub-observer latitude. Right: input (ground truth) compared to the recovered regional spectra. The solid lines represent the average of the input spectra in the recovered region, while the shaded areas indicate the recovered spectra (with 1-$\sigma$ uncertainties). The full dynamic range of the hemisphere-integrated rotational spectroscopic time-series is indicated by the grey shaded region.}
\label{fig:GCM_validation}
\end{figure*}

\subsubsection{Effect of Limb Darkening}
Many discussions of the null space in thermal rotational mapping assume a static surface map and neglect limb darkening. Under these assumptions, all spherical-harmonic modes with odd degree \(\ell>1\) have zero disk-integrated light-curve signature at any inclination. For an exactly equator-on view, the null space is larger and additionally includes modes with even \(\ell\) and odd \(|m|\), which are antisymmetric about the equator \citep{Cowan2013}. However, \cite{Luger2021h2} showed that multiplying the surface map by a limb-darkening profile effectively increases the angular degree of the harmonics and can lift modes from the null space. In Appendix \ref{sec:limb_darkening}, we show that for the JWST observations of SIMP 0136 examined here, including limb darkening significantly alters the inferred surface map. In particular, the observed third temporal Fourier component does not by itself uniquely require north--south asymmetry or a departure from an exactly equator-on viewing geometry, contrary to the inference made by \cite{Akhmetshyn2025}. \citet{Plummer2025} similarly noted that the interpretation of odd temporal harmonics depends on the limb-darkening treatment.

We used a standard quadratic law for the limb darkening
and calculated the wavelength-dependent limb-darkening coefficients for each object using the open-source code \texttt{PICASO} \citep{Batalha2019}. This code can be used to calculate a three-dimensional radiative transfer model assuming a one-dimensional (i.e., spherically symmetric) profile for the brown dwarf. Here, we used the best-fit model from our atmospheric retrieval on the time-averaged observed spectrum for each object.
We then fit the limb-darkening coefficients to the emergent flux at different angles. We plot the resulting limb-darkening coefficients for both objects in Appendix \ref{sec:limb_darkening}.



\subsection{Identifying Spectrally Distinct Regions}
\label{sec:grouping}
After fitting the rotational light curves in each bandpass to derive corresponding brightness maps, we combined these maps to produce a grid of rotation-inferred surface spectra. Our goal is to use this spectral grid to identify the physical mechanisms that drive the observed spectral shape variations across this grid. Since our maps are derived from spherical harmonics, they can be evaluated at arbitrarily fine grid resolutions. We calculated our maps on an equal-area grid with a resolution of roughly $6^{\circ}$ in longitude and latitude. Although the spectra at this grid resolution are highly uncertain and covariant, we mitigated these effects by averaging the spectra over larger regions that have much smaller uncertainties.

Here, we chose to group the rotation-inferred spectra into $3$--$5$ distinct regions based on similarities in their spectral shapes. The underlying assumption is that different physical mechanisms (e.g., cloud radiative feedback, temperature anomalies, chemical disequilibrium) produce distinguishable spectral signatures. \cite{Challener2025} grouped spatially resolved spectra from eclipse mapping with K-means. We instead performed the grouping in principal-component space, projecting the logarithms of the rotation-inferred surface spectra onto orthogonal variability axes. Unlike PCA applied directly to disk-integrated time series \citep[e.g.][]{Akhmetshyn2025,Wang2026,Schrader2026}, our decomposition occurs after the map inversion. The result remains model-dependent because the inferred surface spectra depend on the mapping basis, regularization, geometry, and limb darkening.

For the two objects examined here, we found that the first two principal components (PCs) captured most ($>90\%$) of the variance, allowing us to project our high-dimensional spectral grid onto a two-dimensional plane. We identified spectral end-members by finding the vertices of the spectra in the projected plane (see Appendix \ref{sec:PCA_Analysis}). While the vertices of this projected distribution represent the extrema of the physical parameter space, each one corresponds to a single 6 degree square spatial grid point and therefore carries relatively large uncertainties from the mapping inversion step. We mitigated these uncertainties by averaging over the 100 nearest neighbors of each vertex, which tend to be spatially co-located. We selected the number of nearest neighbors to be large enough that the intrinsic noise of the averaged spectrum is comparable to our modeling uncertainties. We found that our resulting regionally averaged spectra are relatively insensitive to the number of nearest neighbors used in the average or the choice of polygon (number of distinct spectral end-members) within reasonable bounds.

The final rotation-inferred regional spectra exhibit greater contrasts than the disk-integrated phase-resolved spectra. We therefore use a small number of regional end-members as inputs to the atmospheric model rather than fitting every phase spectrum. Greater contrast does not by itself imply greater statistical significance: the regions are covariant outputs of the same inversion, and the separate retrievals below omit that covariance. We use their posterior differences only to describe which atmospheric states can reproduce each end-member under a common model.

\subsection{Validation on General Circulation Model}
\label{sec:validation}
To validate our framework, we applied the full mapping and clustering procedure to synthetic spectroscopic light curves generated from a GCM of VHS~1256b \citep{Tan2025}. We used \texttt{petitRADTRANS} \citep{Mollire2015, Mollire2019, Nasedkin2024} to compute spatially resolved spectra on the GCM grid, adopting the same JWST/NIRSpec wavelength coverage and instrumental setup as the observations. We then calculated the disk-integrated light curves using the assumed viewing geometries and quadratic limb-darkening profiles. We finally injected noise representative of the JWST/NIRSpec data. We assumed the viewing geometries and limb-darkening profiles to be known in both experiments.

In the first test, we introduced three controlled spectral anomalies on top of the global-mean background. In the blue region, we lowered the C/O ratio to 0.3, compared to a background value of 0.7. In the red region, we reduced the silicate (\ce{MgSiO3}) cloud mixing ratio to half the global average. In the purple region, we increased the temperature by 100~K. These chemical and thermal perturbations produce distinct spectral signatures, so a successful recovery should identify both the spatial footprints of the injected regions and their corresponding spectra. We generated spectroscopic light curves assuming the object is observed edge-on. Indeed, the recovered regions closely match the injected regions, and the recovered spectra agree with the input spectra within the inferred uncertainties (see Figure~\ref{fig:GCM_validation}).

In the second test, we used the unmodified GCM to generate spectroscopic light curves for an inclination of $60^{\circ}$. This case is more challenging because the top-of-atmosphere flux pattern contains substantial small-scale structure rather than a small number of idealized regions. Nevertheless, the clustering identifies the dominant large-scale spectral contrast, including the prominent dark region on the western hemisphere. The recovered regional spectra are consistent with the ground-truth spectra averaged over the corresponding recovered regions.

Together, these tests show that our procedure recovers the dominant spatial components and their representative spectra both for artificial maps with sharp boundaries and for maps with realistic atmospheric structure. Importantly, the recovered regional spectra agree well with the ground-truth spectra averaged over the same recovered regions, even when those regions are not internally homogeneous. These validation tests therefore support our use of regionally averaged spectra as the input to the atmospheric retrieval analysis below.

\begin{figure*}[t!]
\centering
\includegraphics[width=\textwidth]{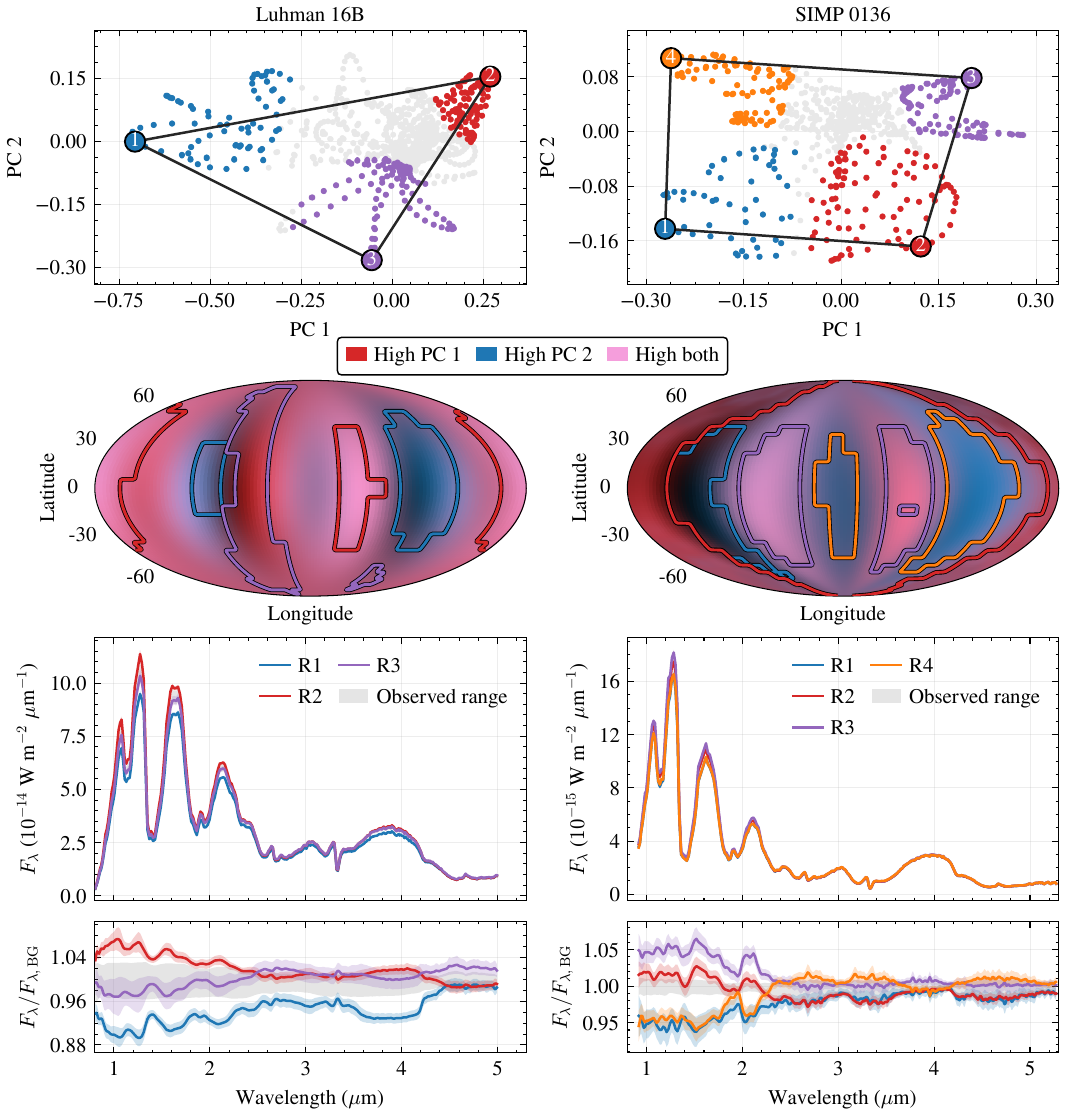}
\caption{Top row: projection of the grid of spectra for Luhman 16B (left) and SIMP 0136 (right) onto the first two principal components. The labels represent the identified extrema points (see Appendix \ref{sec:PCA_Analysis}). Middle row: the corresponding map projections using color to represent the principal component value, with contours marking the identified regions. Bottom row: the rotation-inferred spectra averaged over each region and used as inputs to the atmospheric retrievals; shading denotes the marginal mapping $1\sigma$ uncertainty supplied to those retrievals before uncertainty inflation. Region 1 is blue, Region 2 red, Region 3 purple, and Region 4 orange throughout the paper.}
\label{fig:Luhman16B_S0136_combined}
\end{figure*}

\subsection{Regional Spectra of Luhman 16B and SIMP 0136}
We next applied the same approach to the JWST/NIRSpec time series of Luhman 16B \citep{Biller2024} and SIMP 0136 \citep{McCarthy2025,Nasedkin2025}. For both objects, we marginalized the spherical-harmonic map over degrees $l=2$--6. Luhman 16B is constrained to be close to edge-on \citep{Apai2021}; we therefore marginalized over inclinations of $80^{\circ}$ and $90^{\circ}$. For SIMP 0136, we fixed the inclination at its best-fit value of $80^{\circ}$ \citep{Vos2017}. We verified that our analysis is not sensitive to the assumed inclination. We provide further information on marginalization over uncertain physical parameters, such as inclination and rotational period, in Section \ref{sec:marginalize_physical}.

Both products used wavelength-dependent quadratic limb darkening (see Section \ref{sec:limb_darkening}). We then constructed the posterior distribution of the surface spectra, projected these spectra onto the first two principal components, identified the spectrally distinct regions, and averaged over the regions to define the characteristic regional spectra. We chose to perform the grouping in the first two principal components because they explained more than 98\% of the variance in surface spectra for both targets. In cases where additional components are responsible for higher variance, the algorithm developed here to determine regional end-members naturally extends to $N$ dimensions.

The resulting principal component projections for each object are shown in Figure \ref{fig:Luhman16B_S0136_combined}. We found that the distribution of points resembled a triangle for Luhman 16B, with three distinct vertices. We therefore defined three regional groups (blue, red, and purple) for downstream analysis. For SIMP 0136, the distribution resembled a rectangle with four distinct vertices. We therefore identified four regional groups (blue, red, purple, orange) by drawing a quadrilateral around the points.

We plot the resulting regional spectral groupings in Figure \ref{fig:Luhman16B_S0136_combined}. Both objects are viewed nearly edge-on, so latitudinal features are poorly constrained. These regional spectra should be interpreted as latitudinally weighted longitudinal spectra with little two-dimensional information. In our fits, an equatorial feature is favored over an asymmetric mid-latitudinal feature. Even though both are capable of producing the same light curve, the former requires lower angular power and is preferred under our smoothness prior. The fact that the inferred regions for both Luhman 16B and SIMP 0136 lie near the equator therefore does not definitively establish that they are located there; instead, it means that the data do not require north--south asymmetric structure under the adopted prior.

\begin{figure*}[t!]
\centering
\includegraphics[width=\textwidth]{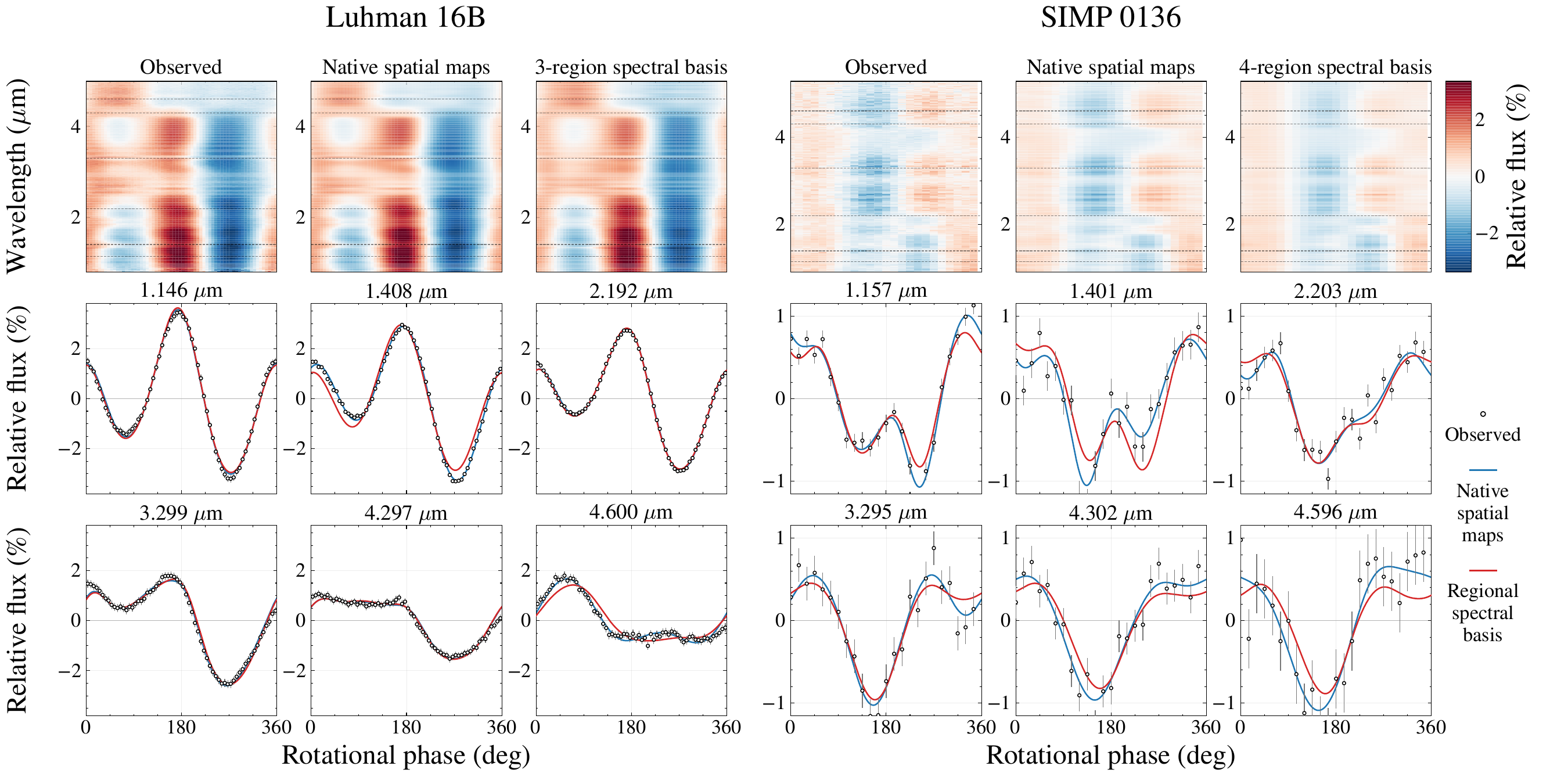}
\caption{Observed and reproduced chromatic variability for Luhman~16B (left three columns) and SIMP~0136 (right three columns). The top row compares the mean-normalized observed wavelength--phase variability with predictions from the native spatial maps and the regional spectral basis; horizontal dashed lines mark the six wavelengths shown in the lower two rows. Black points show the observed light curves, while blue and red curves show the native-map and regional-basis predictions, respectively.}
\label{fig:Luhman16B_S0136_lightcurve_closure}
\end{figure*}

We plot the corresponding regionally averaged spectra in Figure \ref{fig:Luhman16B_S0136_combined}. As expected, we found that the regional spectra for both objects exhibit greater differences than the phase-resolved spectra (grey shading in Figure \ref{fig:Luhman16B_S0136_combined}).
We then assess how well the spatial mapping and regional spectra reproduce the observed variability in Figure~\ref{fig:Luhman16B_S0136_lightcurve_closure}. The ``native spatial maps" are the full wavelength-dependent surface-intensity maps inferred by the original spherical-harmonic inversion. The ``regional spectral basis maps" represent the spectrum at each spatial location in the map as a linear combination of the three (Luhman~16B) or four (SIMP~0136) regional spectra. The predicted light curves are then reconstructed through the same rotational kernel. The close agreement among the observed, native-map, and regional-basis light curves demonstrates that the compact regional basis set recovers the dominant chromatic variability, explaining 96.2\% and 80.5\% of the centered observed variance for Luhman~16B and SIMP~0136, respectively.

\section{Atmospheric Retrieval Setup}
\label{sec:retrieval}
We used nested sampling \citep{Feroz2013} as implemented in \texttt{PyMultiNest} \citep{Buchner2016} to run a suite of atmospheric retrievals on our characteristic regional spectra for each object using \texttt{petitRADTRANS} \citep{Mollire2015, Mollire2019, Nasedkin2024}. We performed independent atmospheric retrievals for the three Luhman~16B regions and four SIMP~0136 regions. Each fit uses the same pressure grid with 50 layers uniformly spaced in log pressure from $10^{-4}$ to $10^{2}$~bar. This is chosen to optimize computational efficiency. We performed additional tests benchmarking the forward model and established that our choice of 50 layers results in a negligible decrease in accuracy compared to more finely sampled models.

For all independent retrievals presented in this paper, we used 1000 live points, sampling efficiency 0.05, and evidence tolerance 0.5. Each retrieval took approximately 1 million likelihood evaluations to converge. For each target, we picked one region and ran an additional test retrieval with 4000 live points and confirmed that this did not change the posterior.

In principle, global parameters such as gravity, radius, and elemental abundances should be shared across regions in a joint retrieval framework. However, such a fit would be computationally intractable with our current retrieval framework. Our atmospheric model has 22 free parameters, which we fit independently to each region. If we make the reasonable assumption that bulk properties, deep adiabats, and iron clouds are shared, a joint fit of $N$ regions would still have 14 (shared) + 8$N$ (regional) parameters. Recall that the number of likelihood evaluations required to sample a parameter space scales exponentially with the number of parameters, and that each likelihood scales linearly with $N$. This means that joint fits quickly become computationally intractable as $N$ increases.

\subsection{Likelihood Function}
For region $k$, let $r_{k,\lambda}$ be the residual at wavelength $\lambda$, and let $\sigma_{k,\lambda}^{2}$ be the corresponding marginal variance from the map inversion. We fit a multiplicative uncertainty scale $\beta_k$ and use the normalized diagonal Gaussian likelihood for the independent retrievals
\begin{equation}
\ln \mathcal{L}_{k} = -\frac{1}{2}\sum_{\lambda}
\left[
\frac{r_{k,\lambda}^{2}}
{\beta_k^2\sigma_{k,\lambda}^2}
+\ln\left(2\pi\beta_k^2\sigma_{k,\lambda}^2\right)
\right],
\end{equation}
This ensures that each fit retains its region's wavelength-dependent marginal uncertainty, but does not account for the off-diagonal covariance with other regions.



\subsection{Temperature-Pressure Profile Informed by Radiative-Convective Equilibrium}
\label{sec:TP}
Our temperature parameterization follows the gradient-based construction of \cite{Zhang2025}. The profile is defined by the reference temperature $T_{\rm ref}$ at 1~bar and seven temperature gradients ($d\ln T/d\ln P$) at $\log_{10}(P/{\rm bar})=2,1,0.5,0,-0.5,-1,$ and $-2$. We interpolate the gradients with a shape-preserving spline, set the gradient to zero at $10^{-3}$~bar, and integrate upward and downward from $T_{\rm ref}$. The profile is isothermal above $10^{-3}$~bar.

We placed a prior on the value of each temperature gradient in our fit based on a pre-computed grid of radiative-convective equilibrium atmospheric models (Sonora Diamondback, \citealt{Morley2024}) to minimize degeneracies and encourage physical consistency as discussed in \cite{Zhang2023}. For Luhman~16B, all seven gradients have a joint truncated-Gaussian prior conditioned on $T_{\rm eff}=1200$~K. Given the previous detection of a temperature inversion for SIMP~0136 \citep{Nasedkin2025}, we use the joint prior conditioned on 1098~K for gradients at 100, 10, $10^{0.5}$, 1, and $10^{-0.5}$~bar, while assigning an independent $\mathrm{TN}(0,0.1;[-0.3,0.3])$ prior to the gradients at 0.1 and 0.01~bar to allow for a stratospheric inversion (TN stands for Truncated Normal).

\begin{figure*}[t]
\centering
\includegraphics[width=\textwidth]{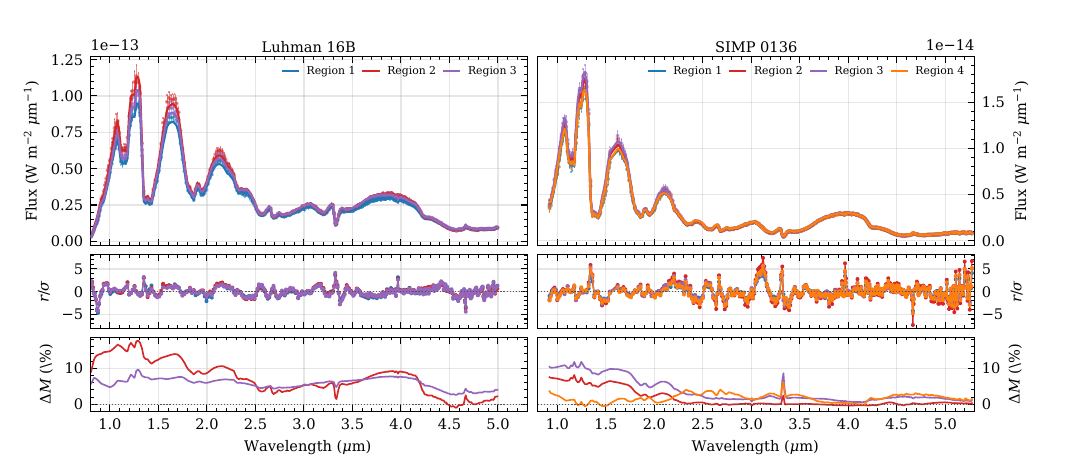}
\caption{Best-fit spectra for the independent Luhman~16B (left) and SIMP~0136 (right) regional fits over the same 0.8--5.3~$\mu$m range. Within each target, the upper panel shows the rotation-inferred regional spectra with scaled marginal uncertainties and the best-fit models, the middle panel shows $(F-M)/(\beta\sigma)$, and the lower panel shows each model relative to Region~1 in percentage difference.}
\label{fig:Luhman16B_bestfit}
\label{fig:S0136_bestfit}
\end{figure*}

\subsection{Chemistry Model}
\label{sec:chemistry}
We used correlated-k opacities for the following species: \ce{Na} \citep{Allard2019_Na}, \ce{K} \citep{Allard2019_K_Allard}, \ce{H2O} \citep{Polyansky2018_H2O_POKAZATEL}, \ce{CH4} \citep{Hargreaves2020_CH4_HITEMP}, \ce{CO} \citep{Rothman2010_CO_HITEMP}, \ce{CO2} \citep{Yurchenko2020_CO2_UCL4000}, \ce{NH3} \citep{Coles2019_NH3_CoYuTe}, \ce{PH3} \citep{SousaSilva2015_PH3_SAlTY}, \ce{H2S} \citep{Azzam2016_H2S_AYT2}, \ce{FeH} \citep{Wende2010_FeH_MoLLIST}, \ce{VO} \citep{Plez2019_VO}, \ce{TiO} \citep{Plez2019_TiO}, and HF \citep{Coxon2015_HX_PotentialFits}. We also accounted for Rayleigh scattering by \ce{H2} \citep{Dalgarno1962_H2_Rayleigh} and He, and collision-induced absorption from \ce{H2}--\ce{H2} and \ce{H2}--He \citep{Richard2012_H2_CIA}.

We allowed the bulk atmospheric C/O ratio and [M/H] to vary as free parameters in our fits. We adopted uniform priors of 0.5 to 1.5 times solar on C/O and $-0.5$ to $0.5$~dex on [M/H]. At each step, we used these elemental abundances to calculate the corresponding equilibrium molecular abundances accounting for rainout. We then allowed these abundances to be modified by transport quenching using a Zahnle--Marley chemical-timescale prescription \citep{Zahnle2014}. We initially parameterized the vertical transport rate as $\log K_{zz,\rm upper}$, $\log K_{zz,\rm lower}$, and the transition pressure as $\log P_{\rm trans}$, following \cite{deRegt2026}. We found that $\log K_{zz,\rm upper}$ was unconstrained in our fits, which favored a constant $\log K_{zz}$ for both targets.

We adopt uniform prior for C/O from 0.5 to 1.5 times solar, and [M/H] from $-0.5$ to $0.5$~dex. Because C/O and [M/H] is a bulk property, differences between independently inferred regional values should be interpreted as regional chemical differences and a sign for more flexible chemistry prescription rather than literal spatial variations of elemental abundances.

In our initial fits, we found that our best-fit model was a poor match to the data in several wavelength ranges. By cross-referencing these residuals with specific opacity cross-sections, we found that removing FeH, VO, PH$_3$, TiO, and HF greatly improved the model evidence ($\Delta \log \mathcal Z > 0$) for both objects. We therefore excluded these species from subsequent fits. The spectral signature of each opacity source can be found in Appendix \ref{sec:supplementary}. Although \cite{Regt2025} detected FeH and HF in Luhman 16B at high spectral resolution in the J-band (1.1--1.4 $\mu$m), this can be explained by the greater sensitivity of high-resolution spectroscopy.

\subsection{Cloud Model}
\label{sec:cloud}
We used the EddySed cloud prescription from \cite{Ackerman2001}. For each cloud species, we included (1) $\sigma_g$, the log-normal width of particle size distribution; (2) $f_{\rm sed}$, the condensate sedimentation efficiency; and (3) $\log \rm{X^{(c)}_0}$, the cloud mass fraction at the cloud base pressure. The same eddy diffusion profile $\rm K_{zz}(P)$ used to calculate quench pressures is also used here. At the relevant temperatures for Luhman 16B and SIMP 0136, silicates and iron are expected to be the dominant cloud species. JWST/NIRSpec observations alone cannot meaningfully constrain the exact mineralogy of the silicate cloud with NIRSpec, and we therefore opted to use \ce{MgSiO3} as the dominant silicate cloud. We tested different options for cloud opacities available in \texttt{petitRADTRANS} and found that amorphous DHS particles are marginally preferred for both objects, but this choice does not affect our retrieved posteriors. We also included an iron (\ce{Fe}) cloud deck in our model, which was required in order to reproduce the observed shape of the highest flux regions of the spectrum.

Our model calculates the locations of the cloud bases and elemental ceilings self-consistently with rainout; the two cloud-base abundances therefore have a joint element-limited prior rather than fixed independent bounds. For example, the iron cloud abundance cannot exceed the iron budget prescribed by the metallicity. Both targets use the same condensate species and cloud parameterization. This means that in total, we have 22 parameters for each regional fit. 
Supplementary Table~\ref{tab:retrieval_supplement} lists the complete parameter set, target-specific priors, posterior ranges, and the derived surface gravities for each region.

\section{Results}

\begin{figure*}[t!]
\centering
\includegraphics[width=\textwidth, trim={0cm 2cm 0cm 0cm}, clip]{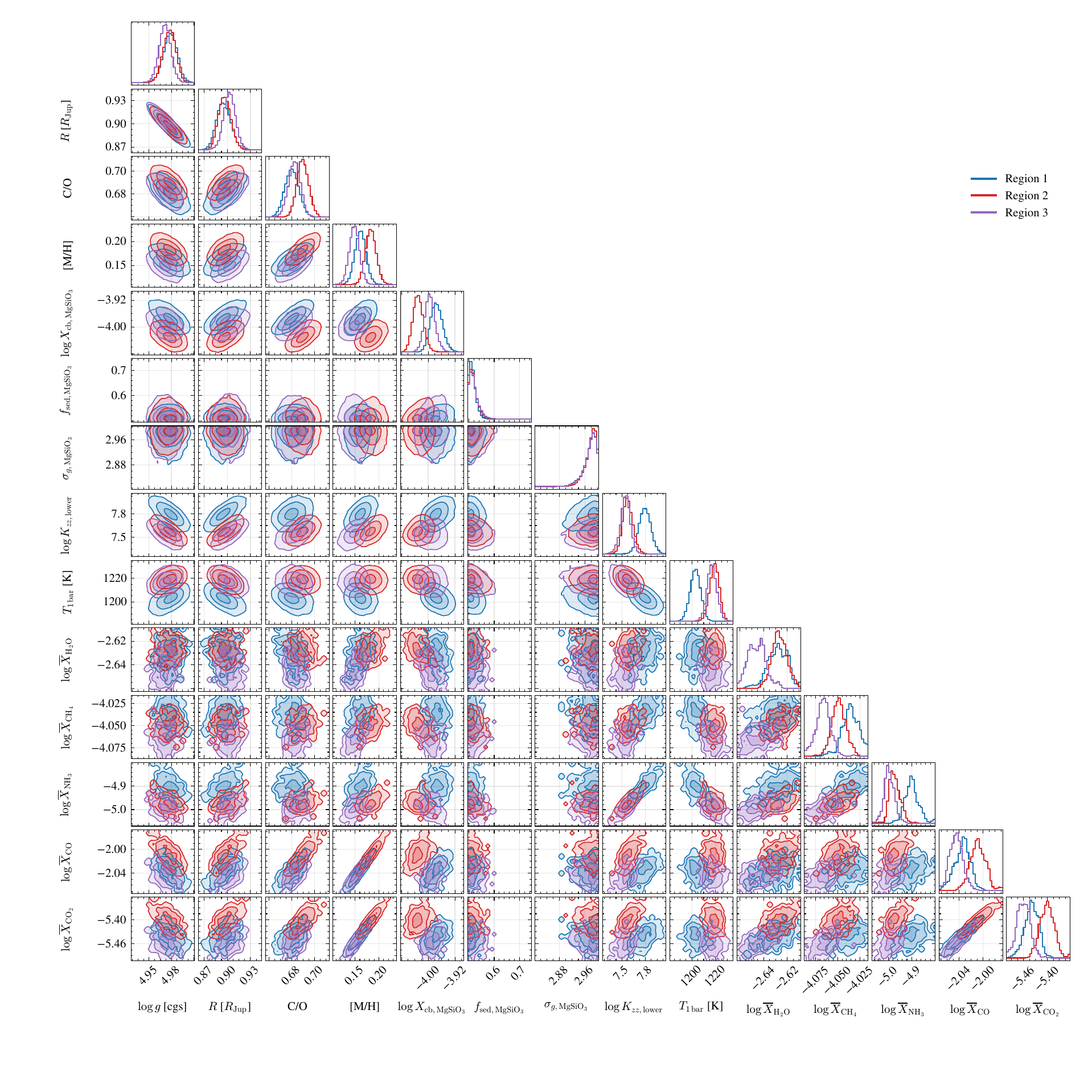}
\caption{Posterior comparison of retrieved parameters and contribution-weighted H$_2$O, CH$_4$, NH$_3$, CO, and CO$_2$ abundances for the three independent Luhman~16B regional retrievals. The full molecular profiles and contribution functions can be found in Appendix \ref{sec:supplementary}. Surface gravity is derived from paired mass and radius samples. Table~\ref{tab:retrieval_supplement} lists the other 14 fitted parameters.}
\label{fig:Luhman16B_independent_overplot}
\end{figure*}

\begin{figure*}[t]
\centering
\includegraphics[width=\textwidth, trim={0cm 2cm 0cm 0cm}, clip]{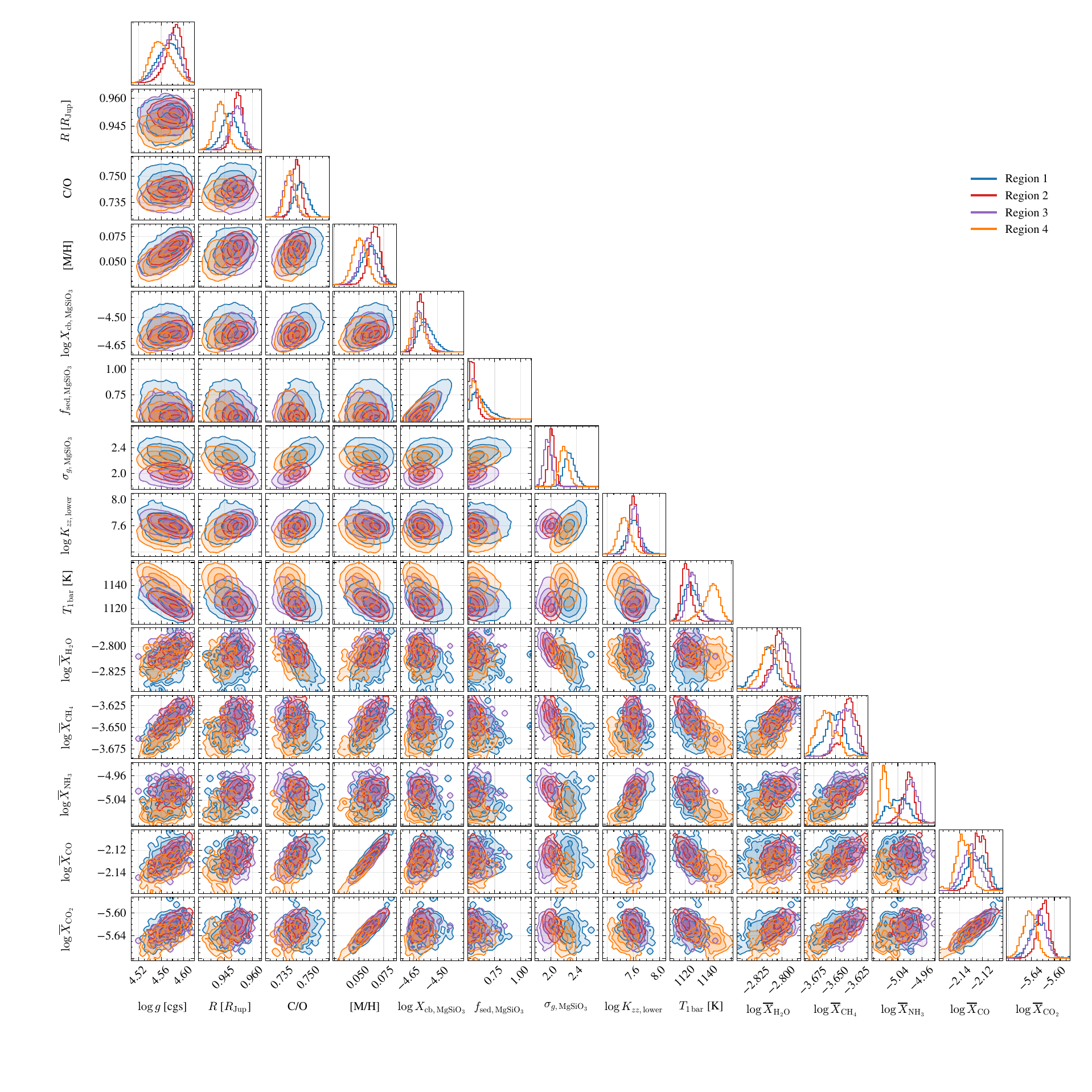}
\caption{Same as Figure \ref{fig:Luhman16B_independent_overplot}, but for the four independent SIMP~0136 regional retrievals.}
\label{fig:S0136_independent_overplot}
\end{figure*}

\begin{figure*}[t]
\centering
\includegraphics[width=1\textwidth]{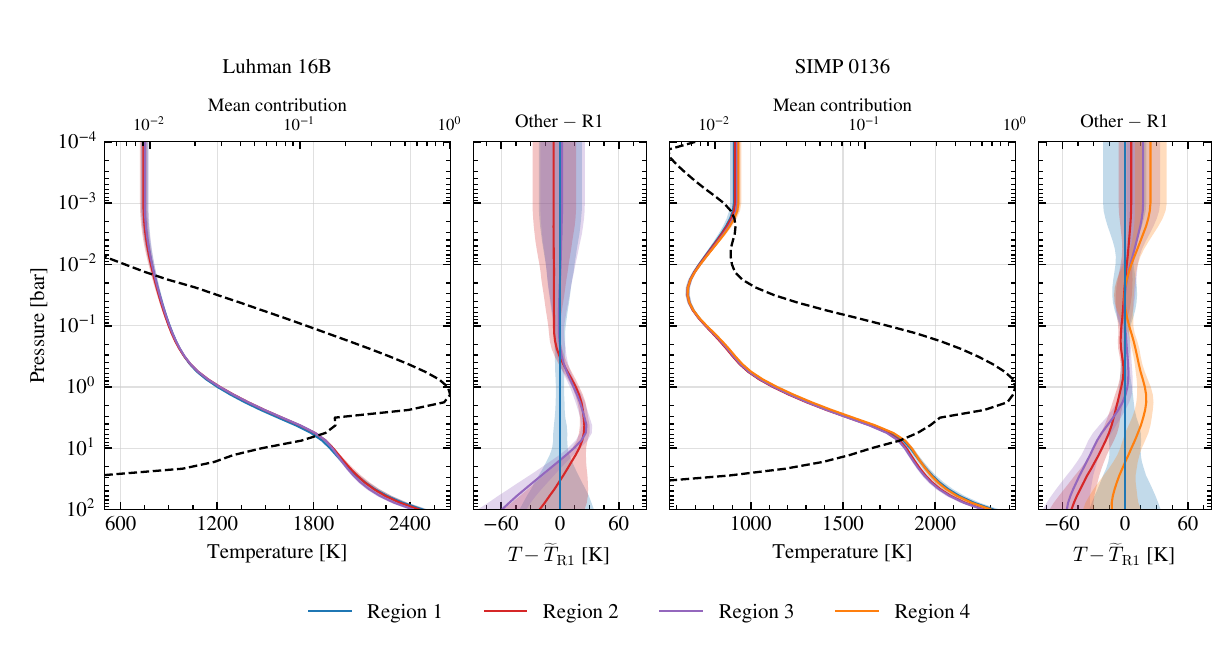}
\caption{Retrieved temperature--pressure profiles and regional temperature differences from the independent regional retrievals. Colored curves and bands show the posterior medians and central 68\% intervals. The narrow panels show $T-\widetilde{T}_{\mathrm{R1}}$, where $\widetilde{T}_{\mathrm{R1}}$ is the Region~1 median. The blue line at zero and its band show Region~1's posterior. The black dashed curve is the region-mean contribution function integrated over wavelength and normalized to unity.}
\label{fig:Luhman16B_S0136_TP}
\includegraphics[width=\textwidth]{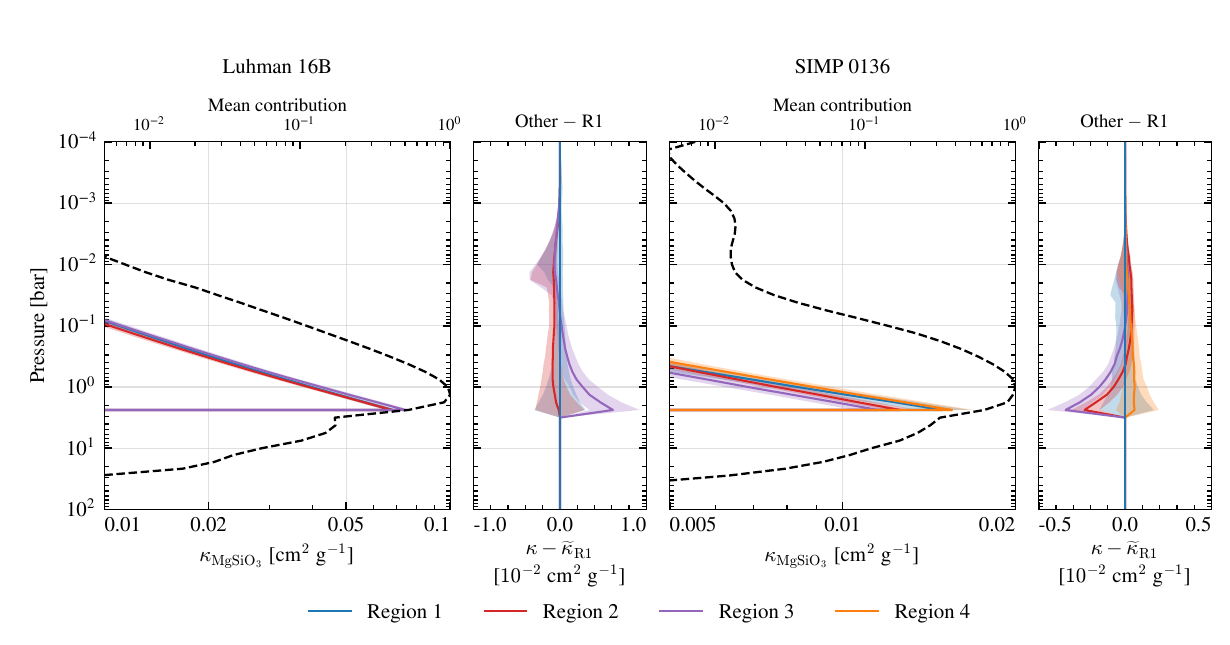}
\caption{Similar to Figure~\ref{fig:Luhman16B_S0136_TP}, but for the \ce{MgSiO3} cloud opacity and its regional differences. The narrow panels subtract the fixed Region~1 median opacity at each pressure, with the same uncertainty convention as Figure~\ref{fig:Luhman16B_S0136_TP}. The opacity is the Planck mean over each target's observed wavelength range of absorption plus $(1-g)$ times the scattering opacity.}
\label{fig:Luhman16B_S0136_cloud_opacity}
\end{figure*}

We plot the best-fit spectra in Figure \ref{fig:Luhman16B_bestfit} and corner plots in Figure \ref{fig:Luhman16B_independent_overplot} and \ref{fig:S0136_independent_overplot}. We find that for both objects, the global parameters appear to be consistent within $1\sigma$ across regions, including surface gravity, radius, C/O, and [M/H]. In the sections below, we discuss the regional differences in cloud, thermal profile, and chemistry revealed by our fits. 

\subsection{Luhman 16B}
We plot the retrieved regional temperature profiles and their relative differences in Figure \ref{fig:Luhman16B_S0136_TP}. We find that Region 1 (blue) is about 25 K cooler than the other regions in the lower photosphere ($1$--$10$~bar). We show the cloud opacity profiles for each region in Figure~\ref{fig:Luhman16B_S0136_cloud_opacity}. The cloud opacity appears to be consistent across the three regions, with Region 3 being the cloudiest. The data favor an iron cloud at higher pressures, but are consistent with models without an iron cloud. All three regions are consistent with a constant $K_{zz}$ profile. Region 1 has the highest retrieved $K_{zz}$, corresponding to the strongest vertical mixing. The silicate-cloud posteriors for all regions accumulate near the lower \(f_{\rm sed}\) and upper \(\sigma_g\) bounds, indicating a global preference for vertically extended clouds with broad particle-size distributions.

We calculate the molecular abundance profiles for each region in Appendix \ref{sec:supplementary}. The overall shape of these profiles is conditional on our adopted chemistry model (equilibrium plus quenching), but the retrieved abundance differences in the photosphere should reflect differences in the shapes of the regional spectra. To identify which chemical differences are constrained by the data and which are driven by our chemistry prescription, we plotted the contribution-weighted abundances derived from posterior samples. We found that \ce{H2O}, \ce{CH4}, and \ce{NH3} do not have strong covariances with other species, whereas the other molecules all co-vary with [M/H]. We therefore conclude that the differences in \ce{CO}, \ce{CO2}, \ce{H2S}, \ce{Na}, and \ce{K} are all driven by differences in [M/H]. Of these molecules, \ce{CO} and \ce{CO2} are the more important absorbers and are likely driving the retrieved metallicity value. We therefore include the posteriors for the abundances of \ce{H2O}, \ce{CH4}, \ce{NH3}, \ce{CO}, and \ce{CO2} in Figure \ref{fig:Luhman16B_independent_overplot}. We find that Region 1 is the most enhanced in \ce{NH3} and \ce{CH4}, in agreement with its lower retrieved photospheric temperature. In contrast, Region 2 appears to have the highest abundance of \ce{CO2} and correspondingly the highest [M/H] and C/O ratio in its photosphere. Region 3 has the lowest abundances of all five molecules, and is uniquely low in its retrieved \ce{H2O} and \ce{CH4} abundances compared to the other two regions.

Returning to the principal-component decomposition
(Figure~\ref{fig:Luhman16B_S0136_combined}), we can now relate PC1 and PC2 to variations in the physical properties of the regions. We find that PC1 separates Regions~1 and~2, which differ in their retrieved lower-photospheric temperatures and vertical mixing. PC2 distinguishes Region~3 from the other two; the posterior for this region favors higher silicate cloud opacity and lower molecular abundances than the other two regions.

\subsection{SIMP 0136}
We find that all four retrieved temperature profiles for SIMP~0136 share a temperature inversion of approximately 250~K (Figure~\ref{fig:Luhman16B_S0136_TP}) at pressures lower than $0.03$ bar. We evaluated the significance of this temperature inversion by calculating the Bayes factor for a fit where the priors excluded inversions, similar to the prior scheme we used for our Luhman~16B fit. We found that the log-evidence values increased by 13.15, 41.12, 23.96, and 27.57 for Regions 1--4 for fits with an inversion. This indicates that the data strongly prefer an inversion. This interpretation is consistent with \citet{Nasedkin2025}, who also inferred a stratospheric inversion on SIMP~0136. As discussed in this study, the observed upper-atmosphere heating might be caused by auroral processes. SIMP~0136 is a known auroral radio emitter, with highly circularly polarized pulses attributed to the electron cyclotron maser instability \citep{Kao2016,Kao2018}. Energetic electrons precipitating along magnetic field lines can deposit energy through collisions with the atmospheric gas, providing a potential source of the heating required to sustain the inversion \citep{Pineda2024}. Although the resulting heating should be localized to regions near the magnetic poles, this object's nearly edge-on viewing geometry means that we can only spatially resolve regions near the equator. We therefore interpret the consistent strength of the temperature inversion across all four regional spectra as evidence for a polar origin (see discussion in Section \ref{sec:best_practices}).

Although the four temperature profiles are all broadly in agreement at low pressures ($<0.1$~bar), we find that they differ notably at higher pressures. Region~4 is warmest in the $0.1$--$10$~bar region, similar to Regions 2 and 3 in our Luhman 16B fit. In contrast, Regions 2 and 3 are notably cooler in their deepest layers (pressures $>1$~bar). We plot the corresponding cloud opacities for these regions in Figure~\ref{fig:Luhman16B_S0136_cloud_opacity}. Region~4 has both the highest temperature and the largest median cloud opacity at pressures near 1 bar, whereas Regions~2 and~3 are both cooler and less cloudy in this same pressure range. This result agrees with predictions for cloud radiative feedback \citep[e.g.][]{Tan2021, Tan2021tdb}: in Region 4, more optically thick clouds absorb more energy from the brown dwarf's interior, resulting in localized warming in the deeper layers relative to regions with less cloud opacity. We see a similar pattern in our Luhman 16B fit, where Region 3 has a higher cloud opacity and warmer temperature between $1$ and $10$~bar than Region 1.

Unlike for Luhman~16B, we find that none of the regions require an iron cloud deck. Similar to Luhman~16B, SIMP~0136's four regional spectra are all consistent with a vertically constant $K_{zz}$; Region~4 has a marginally lower $K_{zz}$. Figure~\ref{fig:S0136_chemistry_overplot} shows the retrieved molecular profiles. Region 4 has the lowest \ce{CH4} and \ce{NH3} abundances, as expected given its higher average temperature.

\section{Discussion}
\subsection{Best Practices for Mapping}\label{sec:best_practices}
The most important practical lesson from rotational mapping is that the recovered map is not unique. As a result, the detailed two-dimensional map depends on the adopted basis, regularization prior, and model complexity. Our evidence-optimized regularization and Bayesian model averaging over \(l_{\rm max}\) do not remove this geometric degeneracy, but they make it explicit: the modes constrained by the light curve inform the regularization scale, while the null-space modes retain prior-dominated uncertainties rather than being silently set to zero.

The same principle applies to physical parameters that enter the mapping operator. The inclination and rotation period are usually inferred from external measurements, and published inclination constraints for variable brown dwarfs are often only accurate to order \(\pm 10^\circ\). Fixing these values can therefore make a map appear more precise than justified by the data. Because our analytic framework makes each design-matrix evaluation inexpensive, we can marginalize over inclination, rotation period, and model complexity in a Bayesian model-averaging framework. This is preferable to selecting a single best geometry and then interpreting the corresponding map as unique.

These considerations are especially important for claims about latitude structure. For both Luhman 16B and SIMP 0136, the nearly equator-on viewing geometry (\(i \sim 80^\circ\)) makes latitudinal features weakly constrained. The sub-observer latitude is best constrained, while uncertainty grows toward the limbs; limb darkening further reduces the contribution from limb regions. A high-latitude feature can therefore project into the light curve similarly to a weaker or smoother feature closer to the sub-observer latitude. In this regime, our identified regions should be interpreted primarily as latitudinally weighted longitudinal end-members, containing little two-dimensional information.

SIMP 0136's rotational light curve is dominated by odd harmonics, which have previously been uniquely associated with north-south asymmetry \citep{Cowan2013, Akhmetshyn2025}. However, as shown in Appendix \ref{sec:limb_darkening}, limb darkening changes the effective basis maps and allows north-south symmetric surface patterns to produce non-zero odd harmonic light curves. Including limb darkening therefore shifts the preferred SIMP 0136 solution toward a more equatorially symmetric map. The broader lesson is that constraints on north-south asymmetry are dependent on the assumed limb-darkening prescription, inclination prior, and surface-map regularization. Rotational mapping is best suited to identifying spectral components from the longitudinal modulation; latitudinal sensitivity should be treated as prior-sensitive unless the viewing geometry and data both strongly favor them. For example, such latitudinal information is present when the secondary eclipses the primary at high impact parameter: the ingress and egress effectively scan the disk of the eclipsed body. This technique is known as eclipse mapping \citep[e.g.][]{Challener2025}, and the framework developed here can be adapted to it.

\subsection{Interpretation and Relation to Recent Work}
For both Luhman~16B and SIMP~0136, we find that the observed differences in the shapes of the regional spectra are influenced by temperature differences in the layers where silicate clouds condense. Although our results are consistent with predictions for cloud radiative feedback, wave-driven circulation can also create coupled horizontal differences in cloud opacity and thermal gradients \citep{Showman2012,Tan2021tdb}. For Luhman~16B, we additionally find evidence for regional chemistry differences that cannot solely be explained by differences in the vertical mixing rate. For SIMP~0136, our ability to infer local chemical variations is limited by the lower signal-to-noise of the JWST spectra.

More significantly, our results challenge the typical two-column framework for fitting the hemisphere-integrated spectra of L-T transition dwarfs, which combines one cloudy column and one clear column with a variable coverage fraction $f$ while keeping the vertical temperature profile constant between the regions. We find that the inferred variations in cloud opacity between regions are relatively small, even though the amplitudes of these regional spectra can differ by as much as 15\%. In our retrievals, these differences are largely driven by small temperature deviations of a few tens of K in the cloud-forming layers, in agreement with \cite{Nasedkin2025}. We note that the regional spectra presented here describe the average properties over $\sim 45^\circ$ in longitude, and it is therefore likely that each regional spectrum is still a mixture of regions with varying levels of cloud opacity. However, we find no need to invoke such a two-column model when fitting our regional spectra. 

Although two-column models can provide a satisfactory fit to individual hemisphere-averaged spectra, they also struggle to explain the observed variability in spectroscopic time-series observations. For example, \cite{deRegt2026} presented a two-column free retrieval on VHS 1256b. Although the quality of fit is excellent, the clear column emission originates from the deep interior and overwhelms the emission from the cloudy column at all wavelengths. This means that a change in the surface fraction of the cloud-free region effectively rescales the amplitude of the summed spectrum, creating the same variability shape across all wavelengths. This fails to explain the wavelength dependence of the observed light curve morphologies for these objects. While \cite{Radcliffe2026} demonstrated that two self-consistent columns with varying $f_{\rm sed}$ can reproduce VHS 1256b's observed wavelength-dependent variability in both molecular absorption and silicate features, their columns diverge in interior temperature due to a lack of heat redistribution in the model.

For SIMP~0136, \cite{Nasedkin2025} ran independent retrievals on each of the 24 phase-resolved NIRSpec spectra. We recovered similar conclusions about the primary drivers of the observed rotational variability with four retrievals rather than twenty-four. \cite{Wang2026} instead focus on the temporal variability patterns. Their method constrains time-varying properties but does not relate them to the spatial domain. Our PCA/end-member step is closest to the approach utilized by \cite{Schrader2026}, who applied PCA to the same observation and mapped the longitudinal contributions of three extreme spectral states. However, our spatial mapping step results in a higher contrast between different spectral components compared to the equivalent end-member spectra derived directly from the spectroscopic time-series data. To demonstrate the benefit of this approach, we ran a set of retrievals on the three spectral end-members identified in \cite{Schrader2026} using the same set-up as before (Figure~\ref{fig:S0136_pca_endmember_retrievals}). We found that the higher spectral contrast in our regional spectra results in more pronounced regional differences than those recovered from the time-series end-members. Given the small rotational variability of SIMP~0136 (about 1\%), this enhanced contrast is essential for meaningfully constraining regional differences in atmospheric properties.

\section{Conclusions}
We developed an analytic rotational-mapping workflow that marginalizes over harmonic complexity and viewing geometry, includes limb darkening, and propagates the map posterior into covariant regional spectra. The recovered regional spectra form a small set of spectral end-members that capture most of the chromatic variability while also quantifying relevant uncertainties in the mapping inversion and assumed viewing geometry.

We report results from independent retrievals on the three regional spectra for Luhman~16B and four regional spectra for SIMP~0136. Although we also explored joint retrievals across all regional spectra, we found that these required significant flexibility in atmospheric properties across regions and were therefore computationally intractable. Our independent fits nonetheless revealed regional differences in the temperature profiles, cloud opacities, and chemistry of these objects. Similar to \cite{Nasedkin2025}, we also identified the presence of a temperature inversion in SIMP~0136. The inversion is recovered from all four regional spectra and is therefore compatible with a phase-persistent auroral contribution from a polar region that remains visible throughout the rotation.


Our framework complements molecular-band light-curve analyses \citep[e.g.,][]{McCarthy2025,Oliveros-Gomez2025}, time-series spectral retrievals \citep{Nasedkin2025}, eigenspectrum retrievals \citep{Wang2026}, and PCA/end-member analyses of time-series spectra \citep{Schrader2026}. By extracting representative regionally averaged surface spectra from the observed rotational variability, we can constrain regional differences in the thermal, chemical, and cloud properties of brown dwarfs. Our results suggest that subtle variations in cloud thickness can drive changes in associated thermal and chemical properties via cloud radiative feedback, but we also identify chemical and thermal variations that appear largely independent of cloud properties.

\section{Data Availability and Acknowledgments}

This work is based in part on observations made with the NASA/ESA/CSA James Webb Space Telescope. We obtained the reduced JWST/NIRSpec time-series spectra of Luhman~16B from Beth Biller \citep{Biller2024} and those of SIMP~0136 from Evert Nasedkin \citep{Nasedkin2025}. The underlying archival data are available from the Mikulski Archive for Space Telescopes (MAST) at the Space Telescope Science Institute, which is operated by the Association of Universities for Research in Astronomy, Inc., under NASA contract NAS5-03127 for JWST. These observations are associated with program GO~2965, \dataset[10.17909/rwee-wr25]{https://doi.org/10.17909/rwee-wr25}, and program GO~3548, \dataset[10.17909/pfnd-md36]{https://doi.org/10.17909/pfnd-md36}, respectively.

AI tools were used for code optimization and visualization of research results in the manuscript. The code and data products produced in this work will be publicly released upon acceptance.

\newpage
\bibliographystyle{aasjournal}
\bibliography{references, additional_references}

\FloatBarrier
\onecolumngrid
\appendix
\section{Supplementary Materials}
\label{sec:supplementary}
\begin{figure}[p]
\centering
\includegraphics[height=0.43\textheight,keepaspectratio]{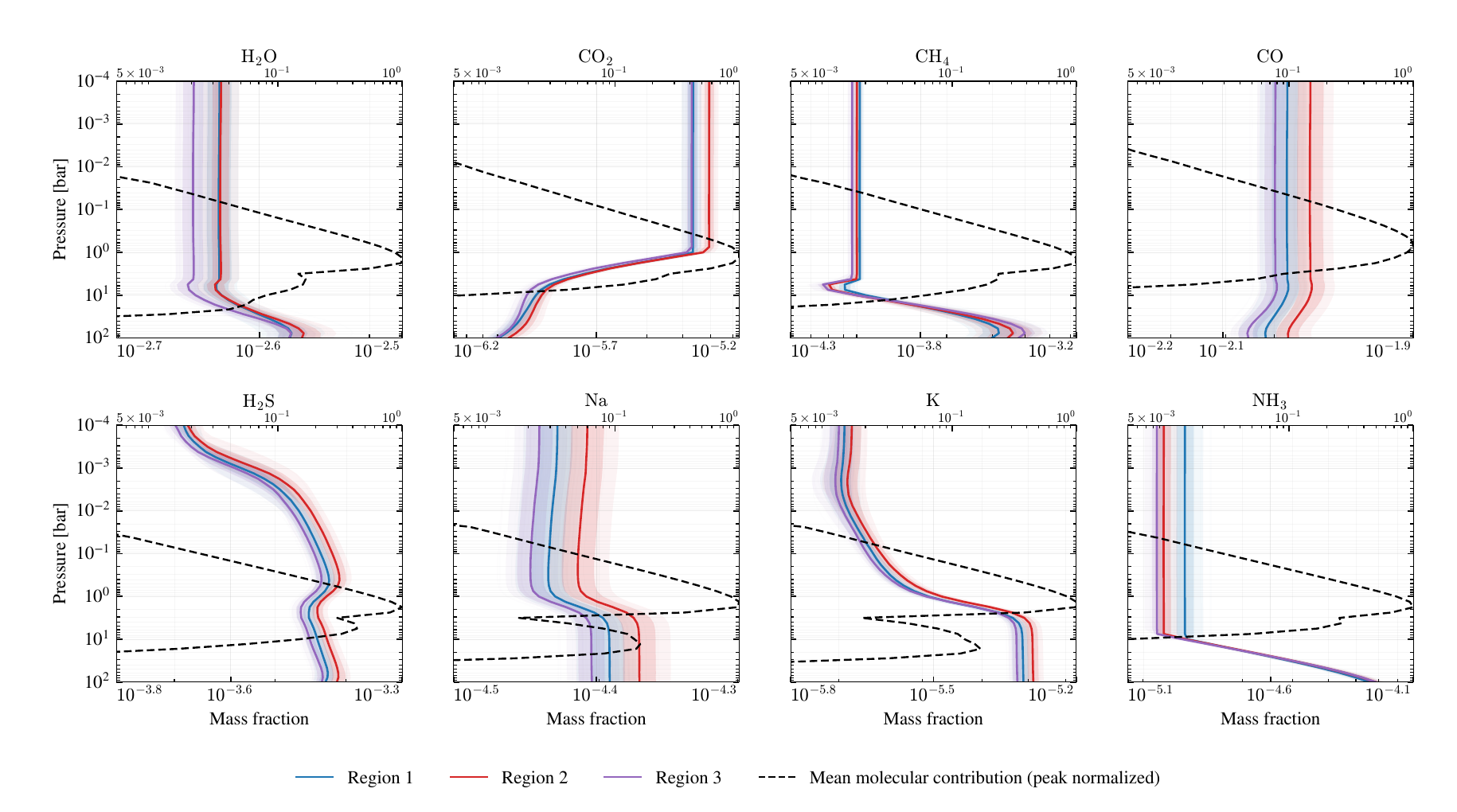}\\[-0.5em]
\includegraphics[height=0.43\textheight,keepaspectratio]{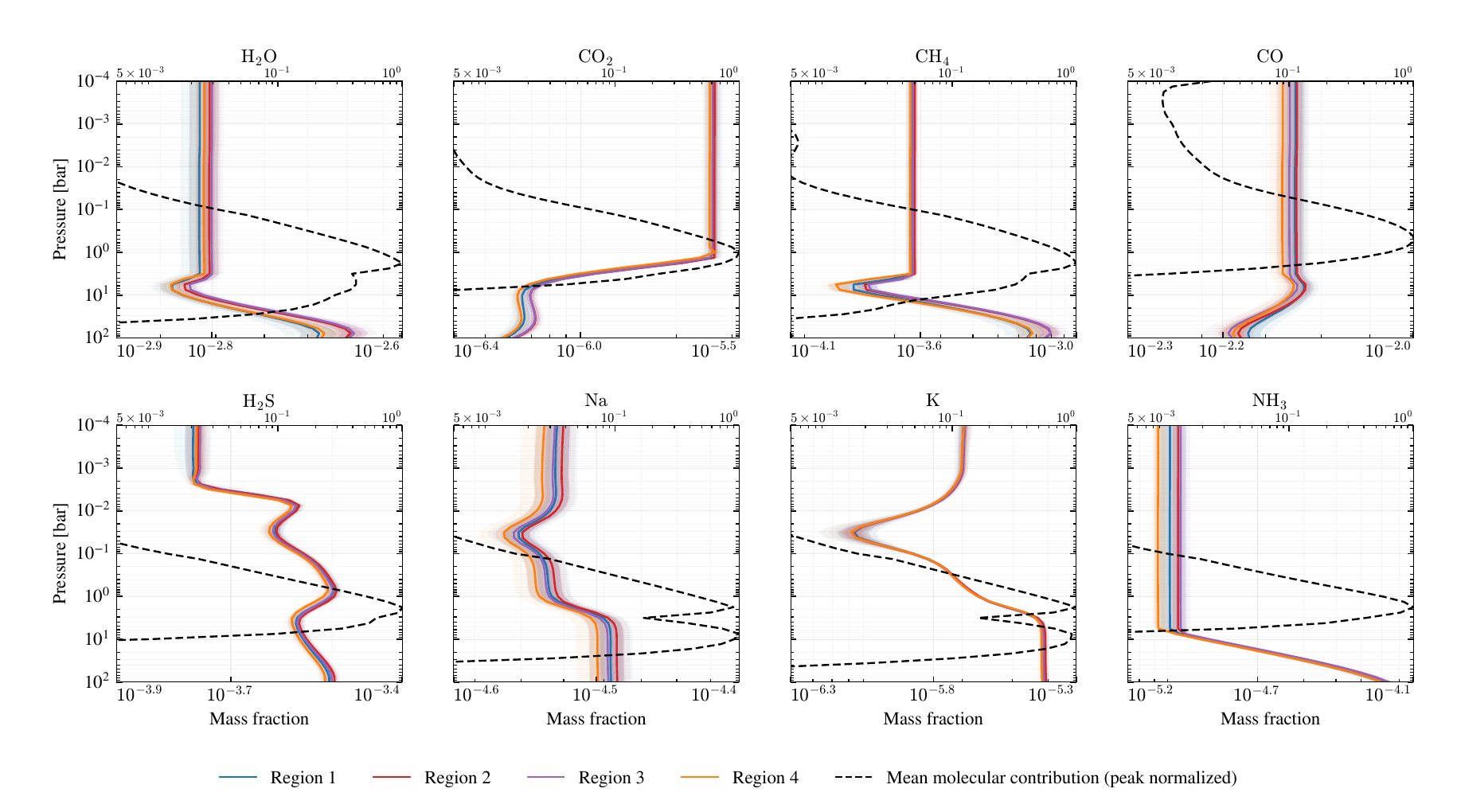}
\caption{Posterior-derived post-rainout, post-quench gas mass-fraction profiles for Luhman~16B (top) and SIMP~0136 (bottom). Colored bands show the central 95\% and 68\% marginal intervals. The black dashed curves show the mean molecular contribution functions, normalized to unity.}
\label{fig:Luhman16B_chemistry_overplot}
\label{fig:S0136_chemistry_overplot}
\end{figure}

\begin{figure}[t]
\centering
\includegraphics[width=0.94\textwidth]{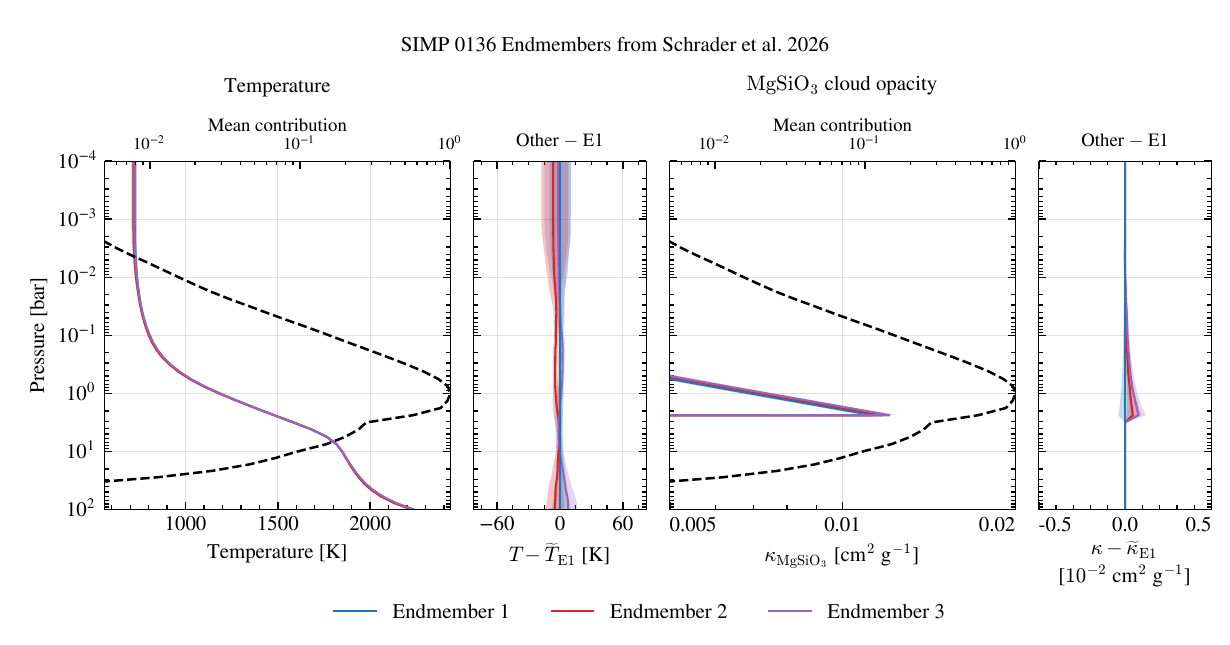}
\caption{Independent retrievals of the three SIMP~0136 spectral end-members identified by \cite{Schrader2026}, using the same atmospheric setup as the regional-spectrum retrievals. The left pair of panels shows the temperature profiles and the E2 and E3 differences relative to E1; the right pair shows the \ce{MgSiO3} cloud opacity and the differences relative to E1. Solid curves and shaded bands denote posterior medians and central 68\% intervals, respectively. Compared with the regional-spectrum retrievals, these time-series end-members exhibit much smaller thermal and cloud-opacity contrasts.}
\label{fig:S0136_pca_endmember_retrievals}
\includegraphics[width=0.92\textwidth]{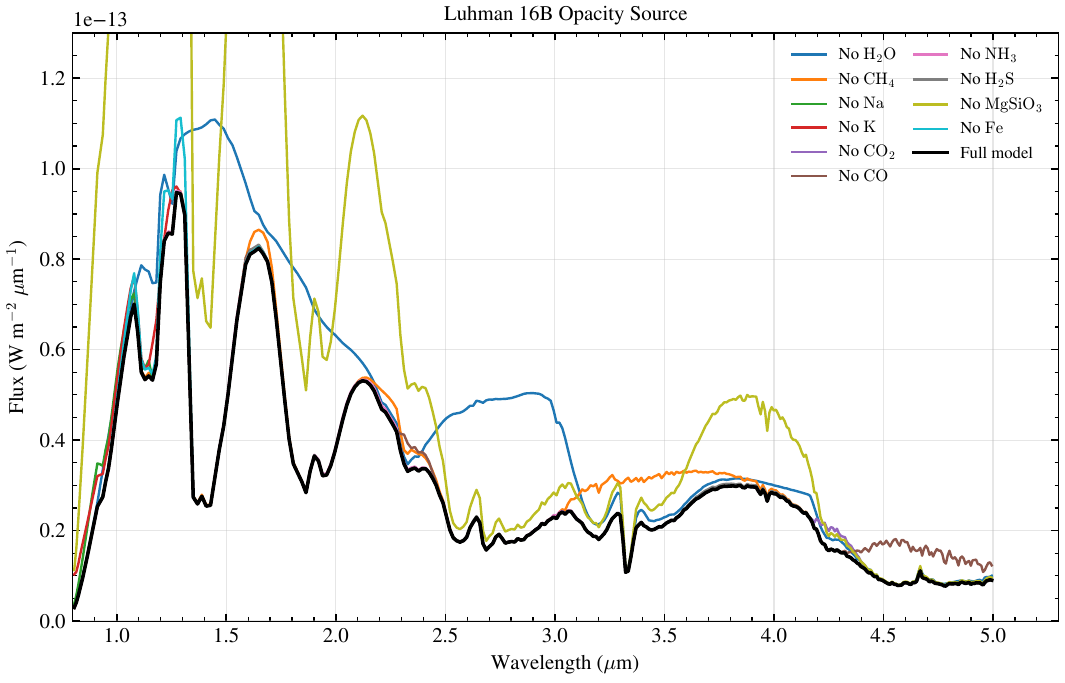}
\caption{Leave-one-opacity-out spectra using the best-fit atmosphere of Luhman~16B Region~1 as an example. The black curve shows the full model. Each colored curve was recomputed after removing one opacity source, with all others held fixed.}
\label{fig:Luhman16B_region1_opacity_ablation}
\end{figure}

Table \ref{tab:retrieval_supplement} lists the parameters, priors, and posteriors of each retrieval. Figure \ref{fig:S0136_chemistry_overplot} plots the retrieved vertical mixing ratios for each molecule. Figure \ref{fig:S0136_pca_endmember_retrievals} presents the retrieval results of the three SIMP 0136 spectral end-members identified by \cite{Schrader2026}. Figure \ref{fig:Luhman16B_region1_opacity_ablation} shows the spectral response to removing individual opacity sources from the Luhman~16B Region~1 best-fit atmosphere.

{
\setlength{\tabcolsep}{2pt}
\renewcommand{\arraystretch}{1.08}
\begin{deluxetable}{l|cccc|ccccc}[h]
\tabletypesize{\scriptsize}
\tablecaption{Target-specific priors and posterior summaries.}
\label{tab:retrieval_supplement}
\tablehead{
\colhead{} & \multicolumn{4}{|c|}{Luhman 16B} & \multicolumn{5}{c}{SIMP 0136} \\
\multicolumn{1}{c|}{Parameter} &
\colhead{Prior} & \colhead{R1} & \colhead{R2} & \multicolumn{1}{c|}{R3} &
\colhead{Prior} & \colhead{R1} & \colhead{R2} & \colhead{R3} & \colhead{R4}
}
\startdata
$\log g$ [cgs] & from $M,R$ & $4.979^{+0.009}_{-0.009}$ & $4.978^{+0.008}_{-0.008}$ & $4.971^{+0.007}_{-0.007}$ & from $M,R$ & $4.570^{+0.017}_{-0.019}$ & $4.584^{+0.010}_{-0.012}$ & $4.574^{+0.015}_{-0.018}$ & $4.554^{+0.017}_{-0.013}$ \\
$M$ [$M_{\rm Jup}$] & $\mathcal{N}(29.4,0.2)$ & $29.43^{+0.16}_{-0.16}$ & $29.44^{+0.16}_{-0.16}$ & $29.44^{+0.17}_{-0.17}$ & $\mathcal{N}(12.7,1.0)$ & $12.95^{+0.50}_{-0.57}$ & $13.48^{+0.32}_{-0.37}$ & $13.18^{+0.44}_{-0.53}$ & $12.30^{+0.48}_{-0.36}$ \\
$R$ [$R_{\rm Jup}$] & $\mathcal{N}(1.0,0.1)$ & $0.895^{+0.009}_{-0.008}$ & $0.896^{+0.008}_{-0.007}$ & $0.903^{+0.007}_{-0.007}$ & $\mathcal{N}(1.22,0.2)$ & $0.950^{+0.004}_{-0.004}$ & $0.954^{+0.002}_{-0.002}$ & $0.954^{+0.003}_{-0.003}$ & $0.944^{+0.003}_{-0.003}$ \\
C/O & $\mathcal{U}(0.5,1.5)\times\odot$ & $0.680^{+0.006}_{-0.006}$ & $0.690^{+0.005}_{-0.005}$ & $0.682^{+0.005}_{-0.005}$ & $\mathcal{U}(0.5,1.5)\times\odot$ & $0.755^{+0.003}_{-0.003}$ & $0.752^{+0.002}_{-0.002}$ & $0.748^{+0.003}_{-0.003}$ & $0.750^{+0.002}_{-0.002}$ \\
{[M/H]} & $\mathcal{U}(-0.5,0.5)$ & $0.161^{+0.011}_{-0.011}$ & $0.183^{+0.011}_{-0.010}$ & $0.148^{+0.010}_{-0.010}$ & $\mathcal{U}(-0.5,0.5)$ & $0.062^{+0.008}_{-0.008}$ & $0.067^{+0.005}_{-0.005}$ & $0.061^{+0.007}_{-0.007}$ & $0.054^{+0.007}_{-0.007}$ \\
$\log X_{\rm cb,MgSiO_3}$ & element-limited & $-3.98^{+0.02}_{-0.02}$ & $-4.03^{+0.02}_{-0.01}$ & $-4.00^{+0.01}_{-0.01}$ & element-limited & $-4.54^{+0.03}_{-0.03}$ & $-4.56^{+0.01}_{-0.01}$ & $-4.56^{+0.02}_{-0.02}$ & $-4.56^{+0.02}_{-0.02}$ \\
$f_{\rm sed,MgSiO_3}$ & $\mathcal{U}(0.5,8)$ & $0.51^{+0.01}_{-0.01}$ & $0.51^{+0.02}_{-0.01}$ & $0.52^{+0.02}_{-0.01}$ & $\mathcal{U}(0.5,8)$ & $0.56^{+0.06}_{-0.04}$ & $0.52^{+0.02}_{-0.01}$ & $0.53^{+0.04}_{-0.02}$ & $0.53^{+0.03}_{-0.02}$ \\
$\sigma_{g,{\rm MgSiO_3}}$ & $\mathcal{U}(1.02,3)$ & $2.98^{+0.01}_{-0.02}$ & $2.98^{+0.01}_{-0.02}$ & $2.98^{+0.01}_{-0.02}$ & $\mathcal{U}(1.02,3)$ & $2.41^{+0.08}_{-0.07}$ & $2.13^{+0.04}_{-0.03}$ & $2.08^{+0.05}_{-0.05}$ & $2.29^{+0.07}_{-0.07}$ \\
$\log X_{\rm cb,Fe}$ & element-limited & $-3.46^{+0.21}_{-0.13}$ & $-3.48^{+0.30}_{-4.62}$ & $-7.03^{+2.28}_{-1.84}$ & element-limited & $-7.57^{+1.47}_{-1.41}$ & $-7.84^{+1.25}_{-1.25}$ & $-7.72^{+1.41}_{-1.36}$ & $-7.77^{+1.28}_{-1.29}$ \\
$f_{\rm sed,Fe}$ & $\mathcal{U}(0.5,8)$ & $4.59^{+1.58}_{-1.16}$ & $4.79^{+1.75}_{-1.56}$ & $4.54^{+2.11}_{-2.25}$ & $\mathcal{U}(0.5,8)$ & $4.34^{+2.21}_{-2.17}$ & $4.51^{+2.02}_{-2.20}$ & $4.68^{+1.99}_{-2.22}$ & $4.35^{+2.12}_{-2.19}$ \\
$\sigma_{g,{\rm Fe}}$ & $\mathcal{U}(1.02,3)$ & $2.21^{+0.51}_{-0.61}$ & $2.03^{+0.60}_{-0.60}$ & $1.95^{+0.62}_{-0.58}$ & $\mathcal{U}(1.02,3)$ & $1.98^{+0.56}_{-0.54}$ & $1.89^{+0.59}_{-0.51}$ & $1.96^{+0.57}_{-0.55}$ & $1.86^{+0.58}_{-0.49}$ \\
$\log K_{zz,{\rm lower}}$ & $\mathcal{U}(3,14)$ & $7.79^{+0.07}_{-0.07}$ & $7.57^{+0.06}_{-0.06}$ & $7.55^{+0.05}_{-0.06}$ & $\mathcal{U}(3,14)$ & $7.70^{+0.10}_{-0.09}$ & $7.67^{+0.05}_{-0.05}$ & $7.72^{+0.07}_{-0.06}$ & $7.46^{+0.10}_{-0.09}$ \\
$\log K_{zz,{\rm upper}}$ & $\mathcal{U}(3,14)$ & $8.34^{+3.33}_{-3.10}$ & $8.28^{+3.33}_{-3.19}$ & $8.34^{+3.46}_{-3.19}$ & $\mathcal{U}(3,14)$ & $8.47^{+3.12}_{-3.08}$ & $8.36^{+3.17}_{-3.03}$ & $7.89^{+3.24}_{-2.81}$ & $8.57^{+3.18}_{-3.16}$ \\
$\log P_{\rm trans}$ [bar] & $\mathcal{U}(-4,2)$ & $-2.52^{+0.90}_{-0.88}$ & $-2.48^{+0.94}_{-0.91}$ & $-2.52^{+0.95}_{-0.91}$ & $\mathcal{U}(-4,2)$ & $-2.37^{+0.97}_{-0.92}$ & $-2.47^{+0.88}_{-0.87}$ & $-2.49^{+0.90}_{-0.87}$ & $-2.46^{+0.90}_{-0.87}$ \\
$\beta$ & $\mathcal{U}(1,4)$ & $2.895^{+0.101}_{-0.091}$ & $3.798^{+0.107}_{-0.113}$ & $2.011^{+0.071}_{-0.064}$ & $\mathcal{U}(1,4)$ & $3.986^{+0.009}_{-0.015}$ & $3.998^{+0.001}_{-0.002}$ & $3.995^{+0.003}_{-0.006}$ & $3.996^{+0.003}_{-0.005}$ \\
$T(1\,{\rm bar})$ [K] & $\mathcal{U}(800,1800)$ & $1203^{+4}_{-4}$ & $1219^{+4}_{-4}$ & $1217^{+4}_{-4}$ & $\mathcal{U}(800,1800)$ & $1127^{+6}_{-6}$ & $1122^{+3}_{-3}$ & $1128^{+5}_{-4}$ & $1152^{+4}_{-6}$ \\
$\nabla_T(10^2\,{\rm bar})$ & conditioned $\mathcal{N}$ & $0.225^{+0.008}_{-0.008}$ & $0.217^{+0.009}_{-0.008}$ & $0.219^{+0.009}_{-0.010}$ & conditioned $\mathcal{N}$ & $0.199^{+0.006}_{-0.005}$ & $0.199^{+0.005}_{-0.004}$ & $0.203^{+0.006}_{-0.004}$ & $0.198^{+0.005}_{-0.004}$ \\
$\nabla_T(10^1\,{\rm bar})$ & conditioned $\mathcal{N}$ & $0.078^{+0.007}_{-0.008}$ & $0.069^{+0.009}_{-0.007}$ & $0.061^{+0.005}_{-0.005}$ & conditioned $\mathcal{N}$ & $0.060^{+0.007}_{-0.008}$ & $0.054^{+0.004}_{-0.004}$ & $0.055^{+0.005}_{-0.005}$ & $0.049^{+0.006}_{-0.006}$ \\
$\nabla_T(10^{0.5}\,{\rm bar})$ & conditioned $\mathcal{N}$ & $0.246^{+0.005}_{-0.005}$ & $0.244^{+0.004}_{-0.004}$ & $0.247^{+0.004}_{-0.004}$ & conditioned $\mathcal{N}$ & $0.282^{+0.006}_{-0.006}$ & $0.283^{+0.004}_{-0.004}$ & $0.272^{+0.004}_{-0.004}$ & $0.275^{+0.004}_{-0.004}$ \\
$\nabla_T(1\,{\rm bar})$ & conditioned $\mathcal{N}$ & $0.221^{+0.006}_{-0.006}$ & $0.228^{+0.006}_{-0.005}$ & $0.229^{+0.006}_{-0.006}$ & conditioned $\mathcal{N}$ & $0.261^{+0.005}_{-0.005}$ & $0.250^{+0.004}_{-0.004}$ & $0.254^{+0.004}_{-0.004}$ & $0.266^{+0.004}_{-0.004}$ \\
$\nabla_T(10^{-0.5}\,{\rm bar})$ & conditioned $\mathcal{N}$ & $0.105^{+0.006}_{-0.007}$ & $0.117^{+0.007}_{-0.007}$ & $0.111^{+0.007}_{-0.007}$ & conditioned $\mathcal{N}$ & $0.136^{+0.006}_{-0.006}$ & $0.150^{+0.005}_{-0.005}$ & $0.141^{+0.005}_{-0.006}$ & $0.146^{+0.005}_{-0.005}$ \\
$\nabla_T(10^{-1}\,{\rm bar})$ & conditioned $\mathcal{N}$ & $0.068^{+0.009}_{-0.009}$ & $0.068^{+0.010}_{-0.009}$ & $0.068^{+0.009}_{-0.010}$ & conditioned $\mathcal{N}$ & $0.075^{+0.007}_{-0.007}$ & $0.069^{+0.005}_{-0.006}$ & $0.071^{+0.007}_{-0.007}$ & $0.071^{+0.006}_{-0.006}$ \\
$\nabla_T(10^{-2}\,{\rm bar})$ & conditioned $\mathcal{N}$ & $0.046^{+0.009}_{-0.009}$ & $0.047^{+0.009}_{-0.010}$ & $0.045^{+0.009}_{-0.010}$ & conditioned $\mathcal{N}$ & $0.040^{+0.008}_{-0.009}$ & $0.015^{+0.008}_{-0.007}$ & $0.030^{+0.008}_{-0.009}$ & $0.027^{+0.009}_{-0.008}$ \\
\enddata
\tablecomments{C/O bounds are multiples of the adopted solar value (0.55); [M/H] is in dex. Surface gravity is derived from the sampled mass and radius. The TP-gradient priors are target-conditioned truncated Gaussians based on evolutionary profiles at 1200~K for Luhman~16B and 1100~K for SIMP~0136. Cloud-base mass fractions have element-limited priors. Posterior entries are weighted medians with 16th/84th-percentile widths. The uncertainty scale parameter multiplies the regional marginal errors.}
\end{deluxetable}
}

\clearpage

\section{Bayesian Formulation of Light Curve Inversion}
\label{sec:Bayesian_Formulation}
The inversion of rotational light curves into surface maps can be formulated as a Bayesian linear regression problem. For each wavelength bin $\lambda$, we denote the observed light curve as
\[
\mathcal{D} = \{t_m, f_m\}_{m=1}^N,
\]
where \(f_m\) is the normalized flux measured at time \(t_m\). We collect these measurements into a flux vector \(\bm{f} \in \mathbb{R}^N\) and model it as
\begin{equation}
\bm{f} = \bm{A}\,\bm{w} + \bm{\varepsilon},
\end{equation}
where \(\bm{A}=\{\phi_d(t_m)\}_{d=1}^D \in \mathbb{R}^{N\times D}\) is the design matrix, \(\bm{w}\) is the vector of spherical-harmonic coefficients, and \(\bm{\varepsilon}\) is observational noise. Here \(D\) denotes only the number of basis functions (model dimension). Recall that \(\bm{A}\) changes with viewing angle through rotation \citep{Luger2019}.

We assume Gaussian noise,
\[
\bm{\varepsilon} \sim \mathcal{N}(\bm{0},\bm{\Sigma}_n),
\]
with \(\bm{\Sigma}_n=\sigma^2\bm{I}_N\) in the simplest homoscedastic case.

In the absence of prior information on the maps, the maximum-likelihood estimator (MLE) yields the surface map that most closely reproduces the observed light curve \citep{Luger2021h2}.

\subsection{Null Space of the Light Curve Inversion Problem}

It is well known that many spherical harmonics produce no imprint on a rotational light curve.
A simple example is a dipole pattern that is antisymmetric about the equator when viewed equator-on: its contributions cancel upon integration over the visible hemisphere.
We refer to such harmonics as belonging to the \emph{null space} of the light-curve operator.

More generally, for rotational light curves, all harmonics with odd \(l > 1\) and \(m < 0\) are in the null space, independent of the viewing geometry.
For an equator-on viewing geometry (inclination \(\approx 90^\circ\)), harmonics with even \(l\) and odd \(m>0\) also lie in the null space; this degeneracy is partially broken when the object is not viewed exactly equator-on.

If a harmonic \(d\) lies in the null space, its associated basis light curve vanishes:
\[
\phi_d(t_m) = 0 \quad \forall m.
\]
Equivalently, the corresponding column of the design matrix \(\bm{A}\) is identically zero, and \(\bm{A}\) will not be full rank. Similarly, if columns of \(\bm{A}\) are linearly dependent, \(\bm{A}\) will also be rank-deficient. The system is then underdetermined, and the usual MLE solution
\[
\hat{\bm{w}} = (\bm{A}^\top \bm{A})^{-1} \bm{A}^\top \bm{f}
\]
is not well-defined.
A standard remedy is to adopt \emph{regularized least squares}, in which one minimizes
\[
\|\bm{f} - \bm{A}\bm{w}\|^2 + \lambda \|\bm{w}\|^2.
\]
This is equivalent to replacing \(\bm{A}^\top \bm{A}\) with \(\bm{A}^\top \bm{A} + \lambda \bm{I}_D\), which is now non-singular, and yields the solution
\[
\hat{\bm{w}} = (\bm{A}^\top \bm{A} + \lambda \bm{I}_D)^{-1} \bm{A}^\top \bm{f}.
\]
In a Bayesian interpretation, this is identical to imposing a zero-mean Gaussian prior on the weights with covariance \(\lambda^{-1} \bm{I}_D\), i.e.,\ penalizing large coefficients.

As emphasized by \citet{Luger2019}, the rotational light-curve inversion problem is highly underdetermined.
An MLE (or weakly regularized) solution typically overfits the data, and all harmonics in the null space are driven toward zero by the regularizer.
However, one can always add any linear combination of null-space harmonics to \(\hat{\bm{w}}\) without altering the light curve, implying a large family of surface maps that fit the data equally well.
Both the inferred coefficients and their associated uncertainties therefore depend sensitively on the assumed regularization.

This can be made explicit by Bayes' theorem:
\[
p(\bm{w} \mid \mathcal{D}, H)
= \frac{p(\mathcal{D} \mid \bm{w}, H)\,p(\bm{w}\mid H)}{p(\mathcal{D}\mid H)},
\]
where \(H\) denotes model assumptions (e.g., basis choice/design matrix construction, prior, and data noise model). This shows that the posterior surface map depends on both the likelihood (data) and the prior (regularization).

Our goal in this work is to identify distinct surface regions and their emergent spectra that are characteristic of the mechanisms driving the observed variability.
Because of the strong degeneracies in the mapping problem, it is essential to propagate uncertainties in the spherical-harmonic coefficients before drawing physical conclusions.
A Bayesian linear regression framework provides a natural extension of regularized least squares, replacing a single “best-fit” map with a full posterior distribution over maps.

\subsection{General Solution to the Bayesian Linear Regression Problem}
Suppose we adopt a Gaussian prior with mean \(\bm{\mu}_0\) and covariance \(\bm{\Sigma}_0\),
\[
\bm{w} \sim \mathcal{N}(\bm{\mu}_0, \bm{\Sigma}_0),
\]
and assume a Gaussian noise covariance \(\bm{\Sigma}_n\) for the data.

Bayes' theorem gives the posterior
\[
p(\bm{w} \mid \mathcal{D}, H) \propto
p(\mathcal{D} \mid \bm{w}, H)\,p(\bm{w}\mid H),
\]
which is again Gaussian with
\begin{align}
\label{eq:posterior}
\bm{\mu}_{\text{post}} &=
\bm{\mu}_0 + \bm{\Sigma}_{\text{post}}\,\bm{A}^\top \bm{\Sigma}_n^{-1}\!\left(\bm{f} - \bm{A}\bm{\mu}_0\right), \\
\bm{\Sigma}_{\text{post}} &= \left(\bm{A}^\top \bm{\Sigma}_n^{-1} \bm{A} + \bm{\Sigma}_0^{-1}\right)^{-1}.
\end{align}
In our context, the posterior mean \(\bm{\mu}_{\text{post}}\) defines our best estimate of the surface map, while \(\bm{\Sigma}_{\text{post}}\) quantifies the uncertainty and covariance between different spherical-harmonic modes.

The marginal likelihood (or evidence) is obtained by integrating over the weights:
\begin{equation}
p(\mathcal{D}\mid H)
= \int p(\mathcal{D}\mid\bm{w},H)\,p(\bm{w}\mid H)\,d\bm{w}
= \mathcal{N}\!\left(\bm{f}\mid\bm{A}\bm{\mu}_0,\,\bm{C}\right),
\quad
\bm{C} = \bm{\Sigma}_n + \bm{A}\bm{\Sigma}_0\bm{A}^\top.
\end{equation}
The corresponding log-evidence is
\begin{equation}
\label{eq:log_evidence}
\log p(\mathcal{D}\mid H)
= -\frac{1}{2}
\left[
(\bm{f} - \bm{A}\bm{\mu}_0)^\top \bm{C}^{-1}(\bm{f} - \bm{A}\bm{\mu}_0)
+ \log|\bm{C}| + N\log(2\pi)
\right].
\end{equation}

\subsection{Prior on the Spherical Harmonic Coefficients}
\citet{mackay1992a} provides a general recipe for Bayesian linear regression with data-driven Gaussian priors. They introduce two hyperparameters, \(\alpha\) and \(\beta\), where \(\alpha\) is the prior precision of the weights and \(\beta\) is the noise precision. These quantities determine the strength of regularization and hence the effective complexity of the inferred model. The hyperparameters are optimized iteratively by evidence maximization. The resulting \((\alpha,\beta)\) can be interpreted as the regularization strengths that yield the best model in an Occam's-razor sense. In the context of light-curve inversion, a larger \(\alpha\) corresponds to a stronger prior preference for small-amplitude maps, while a larger \(\beta\) corresponds to a stronger belief in the accuracy of the data. In practice, one may choose to fix \(\beta\) using the measurement uncertainties from the data-reduction pipeline and optimize only \(\alpha\).

Following the approach of \citet{mackay1992a} and \citet{tipping2001sbl}, the optimal hyperparameters are obtained by maximizing the log evidence (see Equation~\ref{eq:log_evidence}, but now written as an explicit function of \(\alpha\) and \(\beta\)):
\begin{equation}
\log p(\mathcal{D}\mid H, \alpha, \beta)
= -\frac{1}{2}\!\left[
\beta \|\bm{f}-\bm{A}\bm{m}\|^2
+ \alpha\,\bm{m}^\top\bm{m}
+ D\log\alpha
+ N\log\beta
- \log|\bm{C}|
+ N\log(2\pi)
\right],
\label{eq:evidence_alpha_beta}
\end{equation}
where \(\bm{m} = \bm{\mu}_0 + \beta \bm{S} \bm{A}^\top (\bm{f} - \bm{A}\bm{\mu}_0)\) is the posterior mean vector, and \(\bm{C} = \beta^{-1}\bm{I}_N + \alpha^{-1}\bm{A}\bm{A}^\top\) is the marginal covariance of the data under model \(H\).
Since \(\alpha\) and \(\beta\) appear nonlinearly, analytical maximization is not possible.
Instead, fixed-point updates are obtained by differentiation and rearrangement. Such fixed-point iterations converge to the optimal parameters efficiently.

We define the effective number of parameters as
\[
\gamma = \sum_{i=1}^{D} \frac{\lambda_i}{\alpha + \lambda_i}
= D - \alpha\,\mathrm{tr}(\bm{S}),
\]
where \(\{\lambda_i\}\) are the eigenvalues of \(\beta\bm{A}^\top\bm{A}\), and \(\bm{S}=(\alpha\bm{I}_D+\beta\bm{A}^\top\bm{A})^{-1}\) is the posterior covariance of the weights.

Differentiating the log evidence with respect to \(\alpha\) and setting the derivative to zero yields:
\begin{equation}
\alpha^{\text{new}} = \frac{\gamma}{\|\bm{m}\|^2},
\label{eq:alpha_update}
\end{equation}
where \(\bm{m}\) is the posterior mean vector.

Differentiating the log evidence with respect to \(\beta\) gives:
\begin{equation}
\beta^{\text{new}} = \frac{1}{(\sigma^2)^{\text{new}}}
= \frac{N - \gamma}{\|\bm{f} - \bm{A}\bm{m}\|^2},
\label{eq:sigma_update}
\end{equation}
which updates the estimated noise variance in proportion to the residual error, corrected for the number of effectively determined parameters. After convergence, the resulting \(\alpha\) and \(\beta\) can be used to compute the posterior distribution of the weights and the evidence. The expression for the final posterior distribution is given in Equation~\ref{eq:final_posterior}.

\subsection{Marginalization over Physical Parameters}
\label{sec:marginalize_physical}
The design matrix $\bm{A}$ is implicitly dependent on inclination and period, i.e., $\bm{A}$ = $\bm{A}(i, P_{\rm rot})$. Therefore, the solution depends on these parameters as well. So far, we have assumed that inclination $i$ and period $P_{\rm rot}$ are known. If we only have a prior constraint on $i$ and $P_{\rm rot}$ from previous observations, we need to marginalize over these constraints.

The inclination is constrained through the measured projected rotational velocity \((v\sin i)\), together with the rotation period and radius estimates. The projected rotation speed is
\begin{equation}
v \sin i = \frac{2\pi R}{P_{\rm rot}} \sin i.
\label{eq:veq_from_period_radius}
\end{equation}

The radius of a brown dwarf is typically $\sim1$ Jupiter radius, constrained by evolutionary models \citep[e.g.,][]{Marley2021, Morley2024}. Given a measurement of \(v\sin i\), \(i\) is therefore a function of \(P_{\rm rot}\) and \(R\). We can write the marginalized posterior distribution of the map coefficients as

\begin{align}
\label{eq:map_posterior_inclination_period_marginalized}
p(\bm{w}\mid\mathcal{D},H)
&= \iint p\!\left(\bm{w}\mid i,P_{\rm rot},\mathcal{D}_1,H\right)
p\!\left(i,P_{\rm rot} \mid \mathcal{D}_2, \mathcal{D}_3 \right)\,di\,dP_{\rm rot}\\
&= \iint p(\bm{w}\mid i,P_{\rm rot},\mathcal{D}_1,H)\,p(i \mid P_{\rm rot}, \mathcal{D}_2)\, p(P_{\rm rot} \mid \mathcal{D}_3)\,di\,dP_{\rm rot}\\
&= \iiint p(\bm{w}\mid i,P_{\rm rot},\mathcal{D}_1,H)\,p(i \mid P_{\rm rot}, R, \mathcal{D}_2)\,p(P_{\rm rot} \mid \mathcal{D}_3)\,p(R)\,di\,dP_{\rm rot}\,dR
\end{align}
where we denote the spectroscopic time series as \(\mathcal{D}_1\), \(v\sin i\) measurements as \(\mathcal{D}_2\), and period estimates as \(\mathcal{D}_3\). The marginalization over \(i\) and \(P_{\rm rot}\) must be performed numerically using quadrature, but is still far more tractable than MCMC sampling. The resulting posterior distribution is a Gaussian mixture, which can be approximated as a single Gaussian (see Section~\ref{sec:gaussian_mixture_approximation}).

\subsection{Marginalization over Model Complexity}
\label{sec:model_averaging}
When fitting a surface map, we can choose the maximum degree \(l_{\rm max}\) of the spherical harmonics, which determines the number of coefficients \(D\) and hence the model complexity. Traditionally, one would compute the evidence or the Bayesian Information Criterion for each candidate model (e.g., different \(l_{\rm max}\)) and select the optimal one. Thanks to the closed-form nature of our framework, we can now marginalize over models of different complexities using the model evidence. This allows us to account for model uncertainty in our inference of the surface map and its associated spectra, rather than committing to a single “best” model. This is known as Bayesian model averaging, and has been applied in various contexts in astronomy, including atmospheric retrievals \citep[e.g.,][]{Nixon2024}.

We express the model family as \(\{H_k\}\), where each \(H_k\) corresponds to \(l_{\rm max}=k\). We now marginalize over model index using evidence weights, yielding a model-averaged posterior over \(\bm{w}\).
Let \(K\) denote the set of candidate maximum degrees. With a uniform prior over candidate models, \(p(H_k)=1/|K|\), the posterior probability of each model reduces to
\begin{equation}
p(H_k\mid\mathcal{D})
= \frac{p(\mathcal{D}\mid H_k)}{\sum_{j \in K} p(\mathcal{D}\mid H_j)}.
\end{equation}
The model-averaged posterior for the map coefficients is then
\begin{equation}
p(\bm{w}\mid\mathcal{D})
= \sum_{k\in K} p(\bm{w}\mid\mathcal{D},H_k)\,p(H_k\mid\mathcal{D}).
\end{equation}

\subsection{Approximate Gaussian Mixture as a Single Gaussian}
\label{sec:gaussian_mixture_approximation}
In both Sections~\ref{sec:marginalize_physical} and \ref{sec:model_averaging}, the resulting posterior distributions are Gaussian mixtures, not Gaussian. However, we can approximate a Gaussian mixture as a single Gaussian by matching the first two moments. We verified that this approximation is reasonable in our application.

A Gaussian mixture with \(K\) components can be written as
\begin{equation}
p(\bm{w}\mid\mathcal{D})
= \sum_{k=1}^{K} w_k\,\mathcal{N}\!\left(\bm{\mu}_{\text{post}}^{(k)}, \bm{\Sigma}_{\text{post}}^{(k)}\right),
\end{equation}
where \(w_k\) is the weight for the \(k\)-th model, and \(\bm{\mu}_{\text{post}}^{(k)}\) and \(\bm{\Sigma}_{\text{post}}^{(k)}\) are the posterior mean and covariance under model \(H_k\). The Gaussian approximation to this mixture is a single Gaussian distribution with mean \(\bm{\mu}_{GM}\) and covariance \(\bm{\Sigma}_{GM}\), given by
\begin{equation}\mathcal{N}\!\left(\bm{\mu}_{GM}, \bm{\Sigma}_{GM}\right),
\end{equation}
where the mean of the Gaussian approximation is
\begin{equation}
\bm{\mu}_{GM} = \sum_{k=1}^{K} w_k\,\bm{\mu}_{\text{post}}^{(k)},
\end{equation}
and the covariance of the Gaussian approximation is
\begin{equation}
\bm{\Sigma}_{GM} = \sum_{k=1}^{K} w_k\,\left[
\bm{\Sigma}_{\text{post}}^{(k)} + \left(\bm{\mu}_{\text{post}}^{(k)} - \bm{\mu}_{GM}\right)\left(\bm{\mu}_{\text{post}}^{(k)} - \bm{\mu}_{GM}\right)^\top
\right],
\end{equation}



\subsection{Summary}
In the following, we summarize the basic elements of this framework: the choice of prior, the resulting posterior mean and covariance of the coefficients, and the fixed-point updates for the regularization hyperparameters that control the effective model complexity.
\begin{align}
\text{Model:}\quad
\bm{f} &= \bm{A}\bm{w} + \bm{\varepsilon}, \quad
\bm{\varepsilon} \sim \mathcal{N}\!\left(\bm{0}, \beta^{-1} \bm{I}_N\right) \\[4pt]
\text{Prior:} \quad
p(\bm{w}\mid\alpha) &= \mathcal{N}\!\left(\bm{\mu}_0, \alpha^{-1} \bm{I}_D\right) \\[4pt]
\text{Posterior:}\quad
\label{eq:final_posterior}
p(\bm{w}\mid\mathcal{D},H) &= \mathcal{N}\!\left(\bm{m}, \bm{S}\right), \quad \bm{m} = \bm{\mu}_0 + \beta \bm{S} \bm{A}^\top (\bm{f} - \bm{A}\bm{\mu}_0),\quad \bm{S} = \left(\alpha \bm{I}_D + \beta \bm{A}^\top \bm{A}\right)^{-1} \\[4pt]
\text{Effective number of parameters:}\quad
\gamma &= \sum_{i=1}^{D}  \gamma_i = \sum_{i=1}^{D} \frac{\lambda_i}{\alpha + \lambda_i} = D - \alpha\,\mathrm{tr}\!\left(\bm{S}\right),
\quad \{\lambda_i\} = \text{eigenvalues of } \beta \bm{A}^\top \bm{A} \\[4pt]
\text{Fixed-point updates:}\quad
\alpha_{\text{new}} &= \frac{\gamma}{\bm{m}^\top \bm{m}}, \quad \beta_{\text{new}}  = \frac{N - \gamma}{\left\|\bm{f} - \bm{A}\bm{m}\right\|^2}
\end{align}

\section{Rotation-Inferred Regional Spectra from Spherical Harmonic Maps}

Now that we have obtained the posterior distribution of the spherical-harmonic weights, we can define a grid on the spherical surface onto which we project the surface maps. We define the intensity mapping matrix \(\bm{I}_{\rm map} \in \mathbb{R}^{L\times D}\), whose columns are the spherical harmonic functions evaluated on the grid, \(\{Y^l_m(\bm{r}_\ell)\}_{d=1}^D\), where \(\bm{r}_\ell = (\text{latitude}, \text{longitude})\) and \(L\) is the total number of grid points. Then, \(\bm{i} = \bm{I}_{\rm map}\,\bm{w}\) is the surface intensity vector on the grid. In practice, we define the grid using an equal-area Mollweide projection. This ensures that each grid point represents a constant solid angle.

Recall the posterior distribution of \(\bm{w}\):
\[
\bm{w} \sim \mathcal{N}(\bm{m}, \bm{S}).
\]
The surface intensity \(\bm{i}\) is a linear transformation of \(\bm{w}\), and therefore follows a Gaussian distribution:
\begin{equation}
\label{eq:I_posterior}
\bm{i} \sim \mathcal{N}\!\left(\bm{I}_{\rm map}\bm{m}, \,\bm{I}_{\rm map}\,\bm{S}\,\bm{I}_{\rm map}^\top\right).
\end{equation}

We thus obtain a rotation-inferred surface intensity map with propagated uncertainties at each wavelength bin $\lambda$. In the discussion above, we omitted the superscript $\lambda$ for clarity. Repeating this calculation for every wavelength bin yields the matrix of surface spectra $\bm{F} = \{\pi \bm{i}^\lambda\} \in \mathbb{R}^{L \times M}$, where $M$ is the number of wavelength bins and $\pi$ is the geometric factor that converts intensity to flux density for a uniform surface.

\label{sec:PCA_Analysis}
We define the logarithm of the surface-spectra matrix as \(\bm{X} = \log \bm{F}\) and perform a singular value decomposition (SVD) on the mean-subtracted matrix \(\tilde{\bm{X}}\):
\begin{equation}
\tilde{\bm{X}} = \bm{U}\,\bm{\Sigma}\,\bm{V}^\top.
\end{equation}
The columns of \(\bm{V}\) are the principal-component directions in spectral space, and the diagonal elements of \(\bm{\Sigma}\) are the singular values, whose squares give the variance explained by each component. The explained variance ratio of the first few principal components tells us how many dominant components are needed to approximate the data.

We project the matrix \(\tilde{\bm{X}}\) onto the principal component space:
\begin{equation}
\bm{P} = \tilde{\bm{X}}\,\bm{V} = \bm{U}\,\bm{\Sigma},
\end{equation}
where each row of \(\bm{P}\) contains the principal component amplitudes (scores) for a given surface grid point. Compared to grouping in spectral space, this prevents grouping from being dominated by the high-variance direction (e.g., cloud), allowing us to pick up meaningful variations in secondary directions (e.g., chemistry).

The distribution of projected data often exhibits structure and sheds light on how many distinct end-members are present. To identify these end-members, we find the optimal polygon within the convex hull of the projected data $\bm{P}$ that maximizes the number of enclosed points. We then assign the nearest neighbors to each end-member spectrum and take the mean as the final spectral component.
Let $\ell_n$ be the assignment of rotation-inferred surface spectrum $n$. Then $N_k = \sum_{n=1}^{L} \mathbb{I}(\ell_n = k)$ is the number of points within group \(k\). We define the weight matrix \(\bm{W} \in \mathbb{R}^{K \times L}\) by
\[
\bm{W}_{kn} =
\begin{cases}
\dfrac{1}{N_k}, & \text{if } \ell_n = k, \\[4pt]
0,              & \text{otherwise}.
\end{cases}
\]

Recall that \(\bm{X} = \log \bm{F} \) is the logarithm of the surface spectrum. The cluster-mean spectra are then
\[
\bar{\bm{F}} = \bm{W} \bm{F} \in \mathbb{R}^{K \times M}.
\]
The covariance of the cluster-mean spectra at wavelength \(\lambda\) is
\[
\bm{\Sigma}_{\text{cluster}}^{\lambda} = \bm{W} \left(\pi \bm{I}_{\rm map}\bm{S}^{\lambda}\bm{I}_{\rm map}^\top\right) \bm{W}^\top \in \mathbb{R}^{K \times K},
\]

where \(\pi \bm{I}_{\rm map}\bm{S}^{\lambda}\bm{I}_{\rm map}^\top\) is the covariance matrix of the rotation-inferred surface spectra at wavelength index \(\lambda\) (for \(\lambda=1,\dots,M\)) (Equation \ref{eq:I_posterior}). The regional end-member spectra \(\bar{\bm{F}}\) and covariance matrices \(\bm{\Sigma}_{\text{cluster}}\) are the final products that we fit in our retrievals.



\section{Effects of Limb Darkening}
\label{sec:limb_darkening}
\cite{Luger2019} showed how limb darkening affects spherical-harmonic basis functions. A limb-darkening profile can be expressed as a sum of polynomials in \(x\), \(y\), and \(z\), which can in turn be written as a sum of spherical harmonics. Therefore, weighting the surface intensity by a polynomial limb-darkening profile multiplies the surface map by another spherical-harmonic expansion, and the resulting product is itself a sum of spherical harmonics. The angular degree of each basis function is effectively increased by the degree of the limb-darkening profile. Limb darkening is especially important for odd-\(l\) harmonic degrees. Without limb darkening, these harmonics lie in the null space of the rotational light-curve inversion problem because of longitudinal geometric cancellation. This cancellation is broken by limb darkening, lifting some odd-\(l\) harmonics from the null space.

\begin{figure}[p]
\centering
\includegraphics[width=\textwidth]{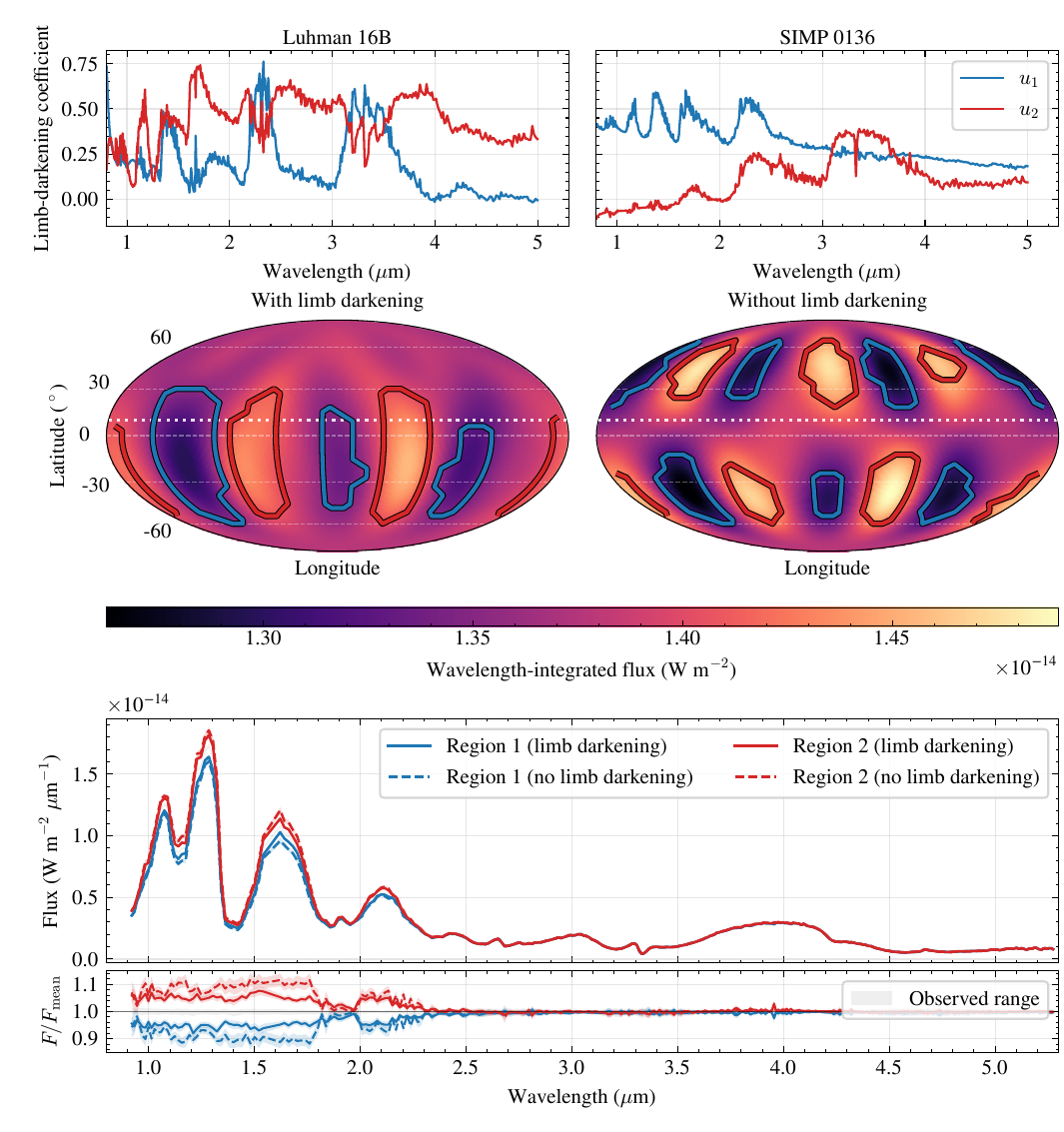}
\caption{Top row: limb-darkening coefficients fitted for Luhman 16B and SIMP 0136. Middle row: wavelength-integrated recovered surface flux for SIMP 0136 with and without limb darkening, with blue and red contours marking the two identified spectral regions and the white dotted line marking the sub-observer latitude. Accounting for limb darkening lifts odd harmonics from the null space and alters the recovered spatial patterns. Bottom row: comparison of rotation-inferred regional spectra with and without limb-darkening corrections, showing that the spectral shapes are similar despite changes in the inferred spatial mapping.}
\label{fig:S0136_limb_darkening_comparison_combined}
\end{figure}

We show the fitted limb-darkening coefficients for Luhman 16B and SIMP 0136 in Figure \ref{fig:S0136_limb_darkening_comparison_combined}. We also confirm that the limb-darkening profiles (intensity as a function of \(\mu\))
are all well described by a quadratic law.
We demonstrate the importance of limb darkening for SIMP 0136, which has a prominent odd harmonic in its NIRSpec rotational light curve, by comparing the rotation-inferred regional spectra with and without limb darkening. The effect of limb darkening on our inferred map is shown in Figure \ref{fig:S0136_limb_darkening_comparison_combined}. Without limb darkening, one light-curve component in SIMP 0136 is fit by the \(l=4,\ m=-2\) harmonic, leading to a recovered map dominated by a checkerboard pattern. After including limb darkening, the \(l=3\) harmonics contribute to the light curve; the limb-darkened \(l=3,\ m=1\) harmonic can reproduce the same component and now dominates the preferred solution.
Despite the large differences in fitted maps, we find that the recovered spectra shown in Figure \ref{fig:S0136_limb_darkening_comparison_combined} have similar shapes in both maps.

We note that in \cite{Cowan2013}, the rotation axis aligns with the spherical-harmonic polar axis in the edge-on case. In \texttt{Starry}'s implementation, the projected disk is represented in sky-plane coordinates \((x,y)\), with the line-of-sight coordinate \(z\), while the default edge-on case has the rotation axis along the sky-projected \(y\)-axis. The two descriptions are related by a rotation of the spherical-harmonic basis. This explains why the light curves derived in \cite{Cowan2013} don't necessarily match with \texttt{Starry}. Similarly, the coefficient labels of the null-space harmonics are different even though the physical null space is the same. For example, the north-south antisymmetric modes are the odd-\(m\) harmonics in \cite{Cowan2013} and the negative-\(m\) harmonics in \texttt{Starry}.

\section{Validation of the Framework on Simulated Data}
\label{sec:GCM_validation}
\begin{figure}[p]
\centering
\includegraphics[width=1\textwidth]{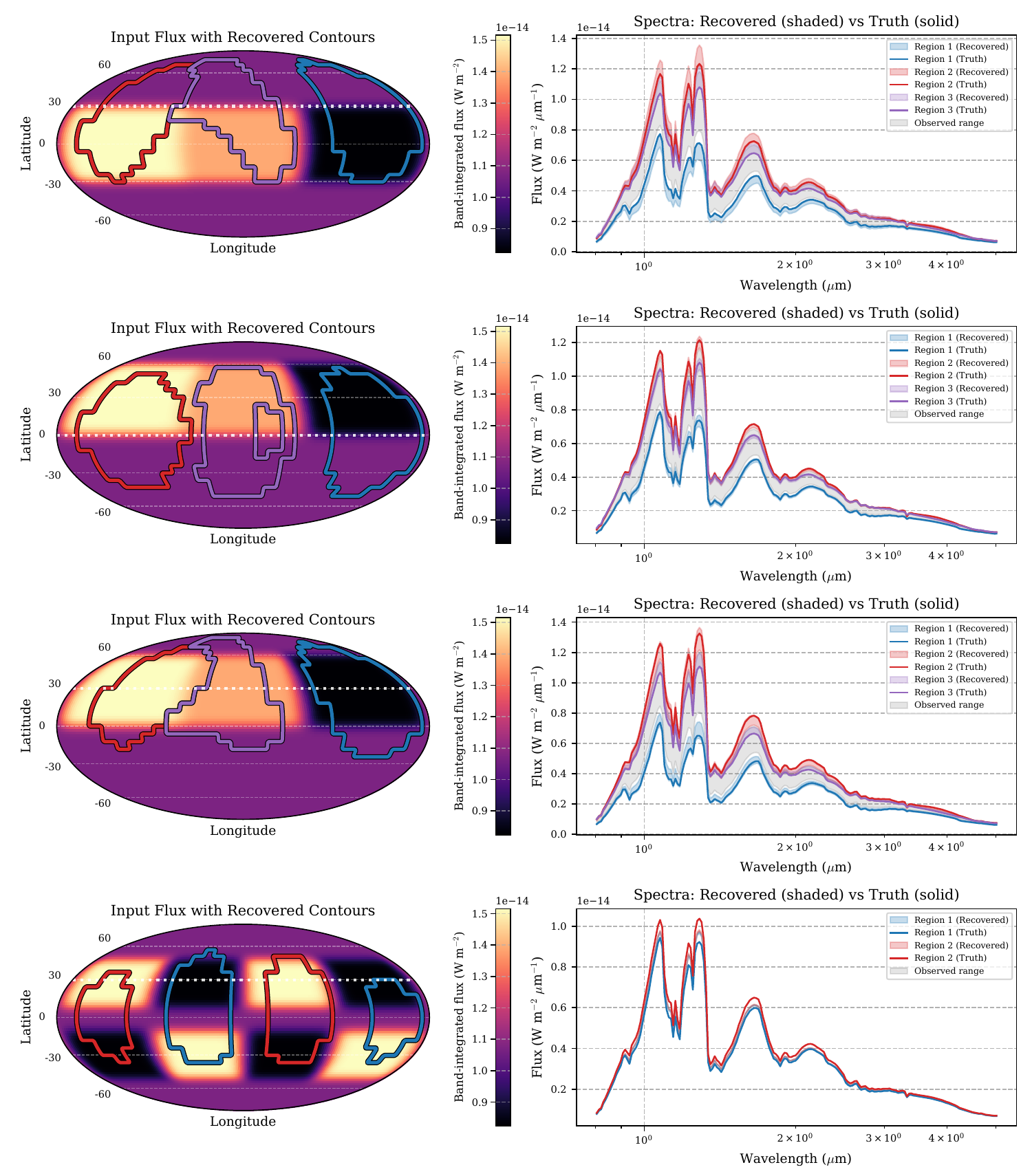}
\caption{Validation tests on simulated GCM light curves with varying feature distributions and viewing geometries. Left column: wavelength-integrated input flux, with colored contours enclosing the recovered spectrally distinct regions; the white dotted line denotes the sub-observer latitude. Right column: recovered regional spectra and their $1\sigma$ uncertainties (shaded) compared with the corresponding input spectra (solid). The gray band spans the full dynamic range of the hemisphere-integrated rotational spectroscopic time series.}
\label{fig:validation_tests}

\end{figure}

In Section \ref{sec:validation}, we tested our framework on cases relevant to Luhman 16B and SIMP 0136 and showed that it recovered both the input regions and their distinct spectra. Here, we expand our test suite to cover a broader range of feature distributions and viewing geometries (see Figure \ref{fig:validation_tests}).

The main conclusions from these tests are:
\begin{enumerate}
    \item For non-edge-on views, the more visible hemisphere is better constrained, and the framework tends to identify regions there even when the input features are symmetric about the equator. However, the recovered spectra still agree with the ground-truth spectra averaged over the identified regions. Thus, the recovered spectra do not necessarily correspond to the most extreme input spectra; instead, they represent average properties of the identified regions.
    \item For edge-on views, north--south asymmetric features are interpreted as north--south symmetric, although their longitudinal distributions are still recovered. The recovered spectra agree well with the ground-truth spectra averaged over the identified regions.
    \item For non-edge-on views, the framework correctly identifies simple north--south asymmetric features.
    \item For non-edge-on views of antisymmetric features, cancellation largely suppresses the variability. The framework then prefers a simpler map with a smaller dynamic range because the prior favors less power in the harmonic power spectrum. In this case, the framework fails to identify the true underlying map, which is unsurprising because the observed spectra exhibit only $\lesssim 2\%$ variability.

    \item An exception to the previous point is that we can identify antisymmetric features of odd spherical-harmonic degree. For non-edge-on views, these maps produce odd temporal harmonics that no even-degree spherical-harmonic map can reproduce. The recovered spectra agree with the ground truth even though cancellation limits the observed variability to $\lesssim 2\%$ (see Figure \ref{fig:GCM_validation}).
\end{enumerate}

\end{document}